\documentclass[12pt]{article}
\usepackage{epsfig}
\usepackage{amsmath}
\usepackage{hhline}
\usepackage{amssymb}
\usepackage{times}
\usepackage{cite}
\usepackage{graphicx}
\usepackage{ifpdf}
\usepackage{color}
\usepackage{colordvi}
\usepackage{epstopdf}
\newlength{\dinwidth}
\newlength{\dinmargin}
\begin{document}
%
%
\def\gsim{\,\lower.25ex\hbox{$\scriptstyle\sim$}\kern-1.30ex%
\raise 0.55ex\hbox{$\scriptstyle >$}\,}
\def\lsim{\,\lower.25ex\hbox{$\scriptstyle\sim$}\kern-1.30ex%
\raise 0.55ex\hbox{$\scriptstyle <$}\,}
\def\bc{\begin{center}}
\def\ec{\end{center}}
\def\bi{\begin{itemize}}
\def\ei{\end{itemize}}
\def\bt{\begin{tabular}}
\def\et{\end{tabular}}
\def\beq{\begin{equation}}
\def\eeq{\end{equation}}
\def\ns{\normalsize}
\def\ul{\underline}
\def\bfm{\boldmath}
\newcommand{\trm}{m_{\perp}}
\newcommand{\trp}{p_{\perp}}
\newcommand{\trmm}{m_{\perp}^2}
\newcommand{\trpp}{p_{\perp}^2}
\newcommand{\sqrts}{$\sqrt{s}$}
\newcommand{\PT}{p_{\perp}}
\newcommand{\xbj}{x}
\newcommand{\gev}{\,\mbox{GeV}}
\newcommand{\GeV}{\mathrm{GeV}}
\newcommand{\GeVS}{\rm GeV^2}
\newcommand{\xp}{x_p}
\newcommand{\xpi}{x_\pi}
\newcommand{\xg}{x_\gamma}
\renewcommand{\deg}{^\circ}
\newcommand{\qsq}{\ensuremath{Q^2} }
\newcommand{\gevsq}{\ensuremath{\mathrm{GeV}^2} }
\newcommand{\rap}{\ensuremath{\eta_{lab}} }
\newcommand{\der}{{\mathrm d}}
\newcommand{\kd}{\textcolor{black}}
\newcommand{\gevc}{\rm GeV}
\def\gs{\mbox{$\gamma^*$}}
\def\Pom{I\!\!P}
\def\Reg{I\!\!R}
\def\xP{x_{I\!\!P}}
\def\aP{\alpha_{\Pom}}
\def\aR{\alpha_{\Reg}}
\def\aPP{\alpha_{\Pom}^{\prime}}
\def\aRP{\alpha_{\Reg}^{\prime}}
\def\api{\alpha_{\pi}(t)}
\def\Rege{\aP(0)-2\api}
\def\as{\alpha_s}
\def\stot{\sigma_{\rm tot}^{\gamma p}}
\def\eprho{e p \to e \rho^0 n \pi^+}
\def\gprho{\gamma p \to \rho^0 n \pi^+}
\def\gpirho{\gamma p \to \rho^0 \pi^+}
\def\sgpr{\sigma_{\gamma p \to \rho^0p}}
\def\gp{\gamma p}
\def\gsp{\gamma^*p}
\def\gpi{\gamma\pi}
\def\gspi{\gamma^*\pi}
\def\Wgp{W_{\gamma p}}
\def\WgpRec{W_{\gamma p,rec}}
\def\Wgpi{W_{\gamma\pi}}
\def\sgsp{\sigma_{\gamma^*p}}
\def\sgspi{\sigma_{\gamma^*\pi}}
\def\sgp{\sigma_{\gamma p}}
\def\sgpi{\sigma_{\gamma\pi}}
\def\sgpiel{\sigma_{\gamma\pi^+ \to \rho^0\pi^+}}
\def\fluxpi{f_{\pi/p}}
\def\fluxpia{\fluxpi(x_L,t)}
\def\fluxg{f_{\gamma/e}}
\def\Gpi{\Gamma_{\pi}}
\def\ptr{p_{T,\rho}}
\def\ptn{p_{T,n}}
\def\dsig{{\rm d}\sigma/{\rm d}}
\def\d2sxp{{\rm d^2}\sgp/{\rm d}x_L{\rm d}\ptn^2}
%
\def\Journal#1#2#3#4{{#1} {\bf #2} (#3) #4}
\def\NCA{Nuovo Cimento }
\def\NIM{Nucl. Instrum. Methods }
\def\NIMA{Nucl. Instrum. Methods {\bf A}}
\def\NPA{Nucl. Phys.   {\bf A}}
\def\NPB{Nucl. Phys.   {\bf B}}
\def\PLB{Phys. Lett.   {\bf B}}
\def\PRL{Phys. Rev. Lett. }
\def\PR{Phys. Rev. }
\def\PRC{Phys. Rev.   {\bf C}}
\def\PRD{Phys. Rev.   {\bf D}}
\def\ZP{Z. Phys. }
\def\ZPA{Z. Phys.     {\bf A}}
\def\ZPC{Z. Phys.     {\bf C}}
\def\EJA{Eur. Phys. J. {\bf A}}
\def\EJC{Eur. Phys. J. {\bf C}}
\def\CPC{Comp. Phys. Commun. }
\begin{titlepage}

\noindent
\begin{flushleft}
{\tt DESY 15-120    \hfill    ISSN 0418-9833} \\
{\tt August 2015}                  \\
\end{flushleft}
\noindent

\vspace{2cm}
\bc
   \begin{Large}

   {\bf \bfm Exclusive $\rho^0$ Meson Photoproduction 
                with a Leading Neutron at HERA}

   \vspace{2cm}
                    H1 Collaboration
   \end{Large}
\ec

\vspace{2cm}
\begin{abstract}
  \noindent \rm
    A first measurement is presented of exclusive photoproduction
    of $\rho^0$ mesons associated with leading neutrons at HERA.
    The data were taken with the H1 detector in the years $2006$ and $2007$
    at a centre-of-mass energy of $\sqrt{s}=319$ GeV 
    and correspond to an integrated luminosity of $1.16$ pb$^{-1}$.
    The $\rho^0$ mesons with transverse momenta $p_T<1$ GeV
    are reconstructed from their decays to charged pions, 
    while leading neutrons carrying a large fraction
    of the incoming proton momentum, $x_L>0.35$, are detected 
    in the Forward Neutron Calorimeter.
    The phase space of the measurement is defined by the photon virtuality
    $Q^2 < 2$ GeV$^2$, the total energy of the photon-proton system
    $20 < \Wgp < 100$ GeV and the polar angle of the leading neutron
    $\theta_n < 0.75$ mrad.
    The cross section of the reaction $\gamma p \to \rho^0 n \pi^+$
    is measured as a function of several variables.
    The data are interpreted in terms of a double peripheral process,
    involving pion exchange at the proton vertex followed by elastic
    photoproduction of a $\rho^0$ meson on the virtual pion. 
    In the framework of one-pion-exchange dominance 
    the elastic cross section of photon-pion  scattering,
    $\sigma^{\rm el}(\gamma\pi^+ \to \rho^0\pi^+)$, is extracted. 
    The value of this cross section indicates significant
    absorptive corrections for the exclusive reaction $\gprho$. 
\end{abstract}

\vspace*{10mm}

\bc
    Accepted by {\it \EJC} 
\ec

\end{titlepage}
\begin{flushleft}
%

V.~Andreev$^{21}$,             
A.~Baghdasaryan$^{33}$,        
K.~Begzsuren$^{30}$,           
A.~Belousov$^{21}$,            
A.~Bolz$^{12}$,                
V.~Boudry$^{24}$,              
G.~Brandt$^{44}$,              
V.~Brisson$^{23}$,             
D.~Britzger$^{10}$,            
A.~Buniatyan$^{2}$
% ...


\bigskip{\it
 $ ^{1}$ I. Physikalisches Institut der RWTH, Aachen, Germany \\
 $ ^{2}$ School of Physics and Astronomy, University of Birmingham,
          Birmingham, UK$^{ b}$ \\
% ...
$ ^{10}$ DESY, Hamburg, Germany \\
% ...
$ ^{12}$ Physikalisches Institut, Universit\"at Heidelberg,
          Heidelberg, Germany$^{ a}$ \\
% ...
$ ^{21}$ Lebedev Physical Institute, Moscow, Russia \\
% ...
$ ^{23}$ LAL, Universit\'e Paris-Sud, CNRS/IN2P3, Orsay, France \\
 $ ^{24}$ LLR, Ecole Polytechnique, CNRS/IN2P3, Palaiseau, France \\
% ...
$ ^{30}$ Institute of Physics and Technology of the Mongolian
          Academy of Sciences, Ulaanbaatar, Mongolia \\
% ...
$ ^{33}$ Yerevan Physics Institute, Yerevan, Armenia \\
% ...
$ ^{44}$ Now at II. Physikalisches Institut, Universit\"at G\"ottingen,
          G\"ottingen, Germany \\
% ...
$ ^a$ Supported by the Bundesministerium f\"ur Bildung und Forschung, FRG,
      under contract numbers 05H09GUF, 05H09VHC, 05H09VHF,  05H16PEA \\
 $ ^b$ Supported by the UK Science and Technology Facilities Council,
      and formerly by the UK Particle Physics and
      Astronomy Research Council \\
% ...
}
\end{flushleft}
\newpage


\section{Introduction}
\label{sec:intro}
Measurements of leading baryon production in high energy particle collisions, 
i.e. the production of protons and neutrons at very small polar angles 
with respect to the initial hadron beam direction (forward direction),
are important inputs for the theoretical understanding of strong interactions 
in the soft, non-perturbative regime.
In $ep$ collisions at HERA,
%
%
a hard scale may be present in such reactions if the photon virtuality, $Q^2$,
is large, or if objects with high transverse momenta, $p_T$,  
are produced in addition to the leading baryon.
In such cases the process usually can be factorised into 
short-distance and long-distance phenomena 
and perturbative QCD often is applicable for the description of 
the hard part of the process.

Previous HERA measurements~\cite{Adloff:1998yg,Chekanov:2002,Chekanov:2003,
Aktas:2004,Chekanov:2005,Chekanov:2007,Aaron:2010ab} 
have demonstrated that in the semi-inclusive reaction $ e+p \to e+n+X$ 
the production of neutrons carrying a large fraction of the proton beam energy
is dominated by the pion exchange process. 
In this picture a virtual photon, emitted from the beam electron, 
interacts with a pion from the proton cloud,
thus giving access to the $\gspi$ cross section and, 
in the deep-inelastic scattering regime, to the pion structure function.

\begin{figure}[t]
\center
 \epsfig{file=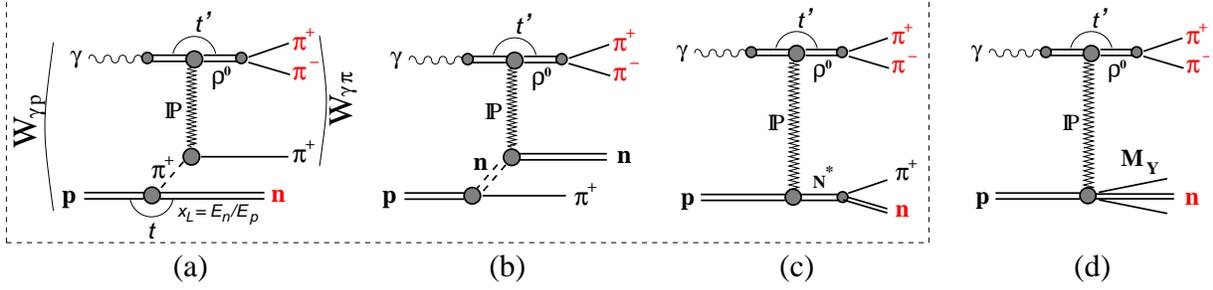,width=\textwidth}
\caption{Generic diagrams for processes contributing to exclusive
         photoproduction of $\rho^0$ mesons associated with leading neutrons at HERA.
         The signal corresponds to the Drell-Hiida-Deck model graphs 
         for the pion exchange (a), neutron exchange (b) and 
         direct pole (c).  
         Diffractive scattering in which a neutron may be produced as a part of the
         proton dissociation system, $M_Y$, contributes as background (d).
         The $N^*$ in (c) denotes both resonant (via $N^+$) and possible non-resonant
         $n+\pi^+$ production.}
\label{fig:FD}
\end{figure}

The aim of the present analysis is to measure exclusive $\rho^0$ production 
on virtual pions in the photoproduction regime at HERA and to extract 
the quasi-elastic $\gamma\pi \to \rho^0\pi$ cross section for the first time. 
Since no hard scale is present, a phenomenological approach, 
such as Regge theory~\cite{regge}, is most appropriate 
to describe the reaction. 
In the Regge framework such events are explained by the diagram 
shown in figure~\ref{fig:FD}a which involves an exchange of two
Regge trajectories in the process $2 \to 3$, known as a 
{\em Double Peripheral Process} (DPP),
or Double-Regge-pole exchange reaction~\cite{DPP}.
This process can also be seen as a proton dissociating into 
$(n,\pi^+)$ system which scatters elastically on the $\rho^0$ 
via the exchange of the Regge trajectory 
with the vacuum quantum numbers, called the ``Pomeron''.

In the past, similar reactions were studied at lower energies in 
nucleon-nucleon and meson-nucleon collisions~\cite{ISR1,ISR2,ISR3,lowE_K,lowE_pi}.
Most of the experimental properties of these reactions were successfully
explained by the generalised Drell-Hiida-Deck model (DHD)~\cite{Deck,Ponomarev,Zotov},
in which in addition to the pion exchange (figure~\ref{fig:FD}a)
two further contributions (figure~\ref{fig:FD}b, \ref{fig:FD}c) are included.
The graphs depicted in figures~\ref{fig:FD}b and \ref{fig:FD}c
give contributions to the total scattering amplitude with similar magnitude 
but opposite sign~\cite{Tsarev,Ponom2}. 
Therefore they largely cancel in most of the phase space, 
in particular at small momentum transfer squared at the proton vertex, $t \to 0$,
such that the pion exchange diagram dominates the cross section~\cite{Zotov}.
One of the specific features observed in these experiments is a characteristic
$t^{\prime}$ dependence at the `elastic' vertex\footnote{In the present analysis 
     elastic vertex corresponds to the $\rho^0\Pom$ vertex, figure~\ref{fig:FD}.}, 
with the slope dependent on the mass 
of the $(n\pi)$ system produced at the other, $pn\pi^+$, vertex, 
and changing in a wide range of approximately $4<b(m)<22$ GeV$^{-2}$.
The Deck model in its original formulation cannot fully describe 
such a strong mass-slope correlation and interference between the amplitudes 
corresponding to the first three graphs in figure~\ref{fig:FD} 
has to be taken into account 
to explain the experimental data~\cite{interf_1,interf_2}.

In the analysis presented here only the two charged pions from the $\rho^0$ decay
and the leading neutron are observed directly.
The pion from the proton vertex is emitted under very small angles
with respect to the proton beam and escapes detection.
This leads to a background contamination from events with a different
final state, which originate from diffractive dissociation 
of the proton into a system $Y$ containing a neutron
(figure~\ref{fig:FD}d). 
Using the H1 detector capabilities in the forward region
such processes can be suppressed to a certain extent.
The residual background contribution is estimated 
from a Monte Carlo model tuned to describe vector meson production 
in diffractive dissociation at HERA. 


The analysis is based on a data sample 
corresponding to an integrated luminosity of $1.16$ pb$^{-1}$
collected with the H1 detector in the years $2006$ and $2007$.
During this period HERA collided positrons and protons with energies of
$E_e=27.6~\GeV$ and $E_p=920~\GeV$, respectively, corresponding to a
centre-of-mass energy of $\sqrt{s}=319~\GeV$.
The photon virtuality is limited to $Q^2<2~\GeV^2$ with 
an average value of $0.04~\GeV^2$.

\section{Cross Sections Definitions}
\label{sec:Kinematics}
The kinematics of the process 
\beq
    e(k) + p(P) \rightarrow e(k') + \rho^0(V) + n(N) + \pi^+ ,
    \label{eq:reaction}
\eeq
%
where the symbols in parentheses denote the four-momenta of the
corresponding particles, is described by the following invariants:
\bi 
 \item the square of the $ep$ centre-of-mass energy $s = (P+k)^2$, 
 \item the modulus of the four-momentum transfer squared at the 
       lepton vertex \\ $Q^2 = -q^2 = -(k-k')^2$,
 \item the inelasticity $y=(q \cdot P)/(k \cdot P)$,
 \item the square of the $\gamma p$ centre-of-mass energy $\Wgp^2 = (q+P)^2 \simeq ys-Q^2$, 
 \item the fraction of the incoming proton beam energy carried by the leading neutron \\ 
       $x_L = (q \cdot N)/(q \cdot P) \simeq E_n/E_p$,
 \item the four-momentum transfer squared at the proton vertex \\
       $t = (P-N)^2 \simeq -\frac{\ptn^2}{x_L} - \frac{(1-x_L)(m_n^2-m_p^2x_L)}{x_L}$,
       and
 \item the four-momentum transfer squared at the photon vertex  $t' = (q-V)^2$.
\ei
Here $E_p, m_p, E_n, m_n$ represent the energy and the mass of the incoming proton
and the outgoing leading neutron, respectively. 
The square of the $\gpi$ centre-of-mass energy 
is then given by $\Wgpi^2 \simeq \Wgp^2(1-x_L)$.

Experimentally, the kinematic variables at the photon vertex 
(the mass $M_{\rho}$, the pseudorapidity $\eta_{\rho}$ and the
transverse momentum squared $\ptr^2$ of the $\rho^0$ meson) 
are determined from the $\rho^0$ decay pions,
while those at the proton vertex ($x_L$ and $\ptn^2$) 
are deduced from the measured
energy and scattering angle of the leading neutron.

In the limit of photoproduction, i.e. $Q^2 \rightarrow 0$,
the beam positron is scattered at small angles and  escapes detection. 
In this regime the  square of the $\gamma p$
centre-of-mass energy can be reconstructed via the variable 
$\WgpRec ^2 = s \, y^{rec}$, where $y^{rec}$ is the
reconstructed inelasticity, measured as
$y^{rec} =  (E_{\rho}-p_{z,\rho})/(2E_e)$. 
Here, $E_{\rho}$ and $p_{z,\rho}$ denote 
the reconstructed energy and the momentum along the proton beam 
direction ($z$-axis) of the $\rho^0$ meson
and $E_e$ is the positron beam energy.
The variable $t'$  
can be estimated from the transverse momentum of the $\rho^0$ meson 
in the laboratory frame via the observable $t'_{rec} = -\ptr^2$ 
to a very good approximation\footnote{A correction accounting 
    for the small, but non-zero $Q^2$ values is applied,
    based on the Monte Carlo generator information, as explained
    in section~\ref{sec:mc}.}.

The cross section of the exclusive reaction~(\ref{eq:reaction}) 
can be expressed as a product of a virtual photon flux $\fluxg$ and 
a photon-proton cross section $\sgp$:
%
\beq
  \frac{{\rm d}^2\sigma_{ep}}{{\rm d}y {\rm d}Q^2} = \fluxg(y,Q^2) \sgp(\Wgp(y)).
  \label{eq:ep2gp}
\eeq 
%
In the Vector Dominance model (VDM)~\cite{Sakurai,Bauer}
%
%
 taking into account both transversely and longitudinally polarised virtual photons
%
%
%
 the effective photon flux is given by
%
\beq
  \fluxg(y,Q^2) = \frac{\alpha}{2\pi Q^2 y}
                     \left\{\left [
                             1+(1-y)^2 - 2(1-y)
                                \left(
                                  \frac{Q^2_{\rm min}}{Q^2}-\frac{Q^2}{M^2_{\rho}}
                                \right)
                             \right ]
                          \frac{1}{\left( 1+\frac{Q^2}{M^2_{\rho}}\right)^2}
                     \right\},
  \label{eq:gflux}
\eeq
%
where $\alpha$ is the fine structure constant and $Q_{\min}^2=m_e^2 y^2/(1-y)$,
with $m_e$ being the mass of the electron 
and $M_{\rho}$ is the $\rho^0$ meson mass.

In the one-pion-exchange (OPE) approximation~\cite{OPE}, which is valid for 
very small $\ptn^2 \sim m_{\pi}^2$, the photon-proton cross section can be 
further decomposed into a pion flux, describing $p \to n\pi^+$ splitting, 
convoluted with a photon-pion cross section:
%
\beq
  \frac{{\rm d}^2\sgp(\Wgp,x_L,t)}{{\rm d}x_L{\rm d}t} = \fluxpia \, \sgpi(\Wgpi). 
  \label{eq:gp2gpi}
\eeq
%
A generic expression for the pion flux factor can be written as follows:  
%
\beq
  \fluxpia = \frac{1}{2\pi} \frac{g_{p\pi n}^2}{4\pi}
                   (1-x_L)^{\Rege}\frac{-t}{(m_{\pi}^2-t)^2} F^2(t,x_L) ,
  \label{eq:piflux}
\eeq
%
where $\aP(0)$ is the Pomeron intercept, 
$\api=\alpha^{\prime}_{\pi}(t-m_{\pi}^2)$ is the pion trajectory,
$g_{p\pi n}^2/4\pi$ is the $p\pi n$ coupling constant
known from phenomenological analysis~\cite{gpiPN} of low energy data,
and $F(t,x_L)$ is a form factor accounting for off mass-shell corrections 
and normalised to unity at the pion pole, $F(m_{\pi}^2,x_L)=1$.
There exists a variety of models for the exact form of the pion 
flux~\cite{flux_Bish,flux_Holt,flux_PSI,flux_KPP,flux_MST,flux_FMS,flux_NSSS}
which typically leads to a $\sim\!30\%$ spread in the predicted 
cross section according to equation~(\ref{eq:gp2gpi}).
Most of models use a non-Reggeized version of equation~(\ref{eq:piflux}), 
i.e. $\aP(0)=1$ and $\api=0$.

\section{Experimental Procedure and Data Analysis}
\subsection{H1 detector}
\label{sec:detector}

A detailed description of the H1 detector can be found elsewhere~\cite{H1detector}.
Only those components relevant for the present analysis are described here.
The origin of the right-handed
H1 coordinate system is the nominal $ep$ interaction point. 
The direction of the proton beam defines  the positive $z$--axis; 
the polar angle $\theta$ is measured with respect to  this axis. 
Transverse momenta are measured in the $x$--$y$ plane.
The pseudorapidity is defined by  $\eta =  -\ln{[\tan(\theta/2)]}$ 
and is measured in the laboratory frame.

The central region of the detector is equipped with a tracking system. 
It included a set  of two large coaxial cylindrical drift chambers (CJC), 
interleaved by a $z$ chamber,  and the central silicon tracker (CST)~\cite{CST}
operated in a solenoidal magnetic field of $1.16~{\rm T}$.
This provides a measurement of the transverse momentum of charged particles with 
resolution 
           $\sigma(p_T) / p_T \simeq 0.002 \ p_T \oplus 0.015$ 
                    ($p_T$ measured in~\gevc), 
for particles emitted from the nominal interaction point with polar angle 
$ 20^{\circ} \leq \theta \leq 160^{\circ}$. 
The interaction vertex is reconstructed from the tracks.
The five central inner proportional chambers (CIP)~\cite{CIP} are
located between the inner CJC and the CST.
%
%
The CIP has an angular acceptance in the range $10^{\circ}<\theta<170^{\circ}$.
The forward tracking detector is used to supplement track reconstruction 
in the region $7\deg<\theta<30\deg$
and improves the hadronic final state 
reconstruction of forward going low momentum particles.

The tracking system is surrounded by a finely segmented
liquid argon (LAr) calorimeter, which covers the polar angle range
\mbox{$4\deg<\theta<154\deg$} with full azimuthal acceptance.  
The LAr calorimeter
is used to measure the scattered electron and 
to reconstruct the energy of the hadronic final state. 
%
%
The backward region ($153\deg<\theta<177.8\deg$) is covered by a
lead/scintillating-fibre calorimeter (SpaCal)~\cite{SpaCal};
its main purpose is the detection of scattered positrons.  

A set of ``forward detectors" is sensitive to the energy flow close to the 
outgoing proton beam direction. It consists of the forward muon detector (FMD),
the Plug calorimeter and the forward tagging system (FTS).
The lead--scintillator Plug calorimeter enables energy measurements to be made 
in the pseudorapidity range $3.5 < \eta < 5.5$.
It is positioned around the beam-pipe at $z = 4.9$~m.
The FMD is a system of six drift chambers which are grouped into two
three-layer sections separated by a toroidal magnet.
Although the nominal coverage of the FMD is
$1.9 < \eta < 3.7$, particles with pseudorapidity up to $\eta \simeq 6.5$ 
can be detected indirectly through their interactions with 
the beam transport system and detector support structures. 
The very forward region, $6.0 < \eta < 7.5$,
is covered by an FTS station which is used in this analysis.
It consists of scintillator detectors surrounding the beam pipe at $z=28$ m.
The forward detectors together with the LAr calorimeter are used here 
to suppress inelastic and proton dissociative background 
by requiring a large rapidity gap (LRG) void of activity 
between the leading neutron and the pions from the $\rho^0$ decay. 

\begin{figure}[ppp]
\center
 \epsfig{file=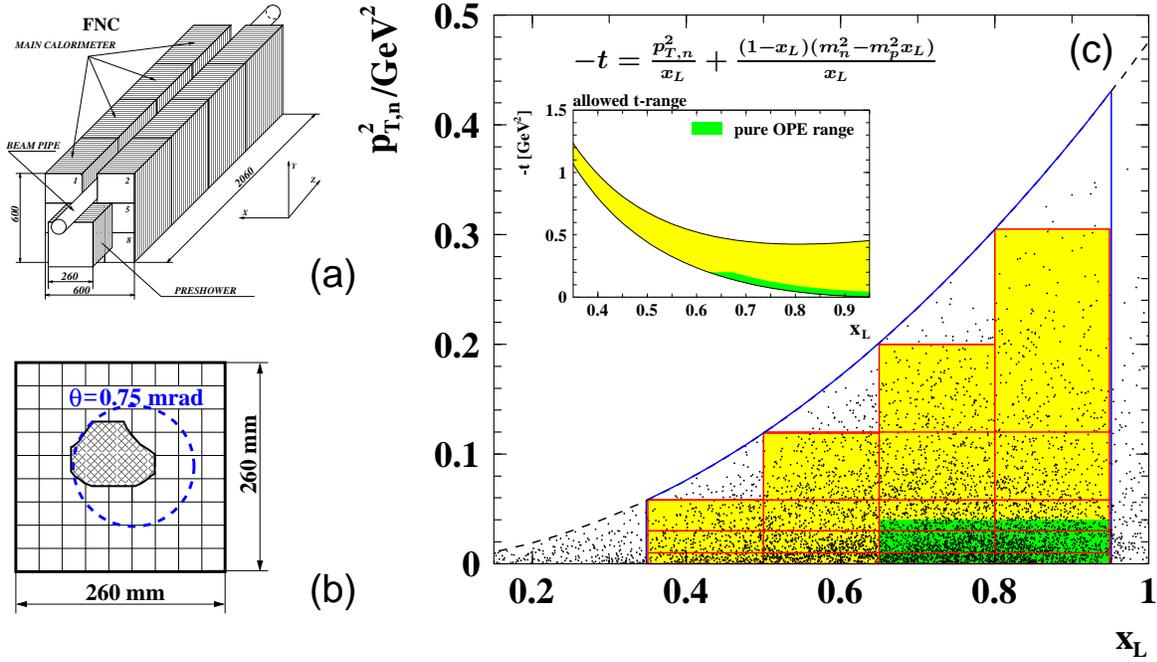,width=\textwidth}
 \setlength{\unitlength}{1cm}
\caption{Sketch of the Forward Neutron Calorimeter (a), 
         acceptance in the azimuthal plane (b) and the $(x_L,\ptn^2)$ plane (c). 
         The shaded area in figure (b) is the projected aperture 
         limited by the proton beamline elements.
         The insert of figure (c) shows
         the acceptance in terms of the $x_L$ and $t$ variables.
         The dark green area indicates the OPE2 region, 
         $\ptn<0.2$ GeV and $x_L>0.65$.
         The curve in the main figure corresponds to the angular cut 
         $\theta_n<0.75$ mrad, and the grid shows 
         the binning scheme used for the double differential
         cross section measurement, $\d2sxp$.
         The dots are events from the preselection sample described
         in the leftmost column of table~\ref{tab:selection}. } 
\label{fig:FNC}
\end{figure}

Neutral particles produced at very small polar angles 
can be detected 
in the forward neutron calorimeter (FNC)~\cite{Aaron:2010ab,FNC}, 
which is situated at $106$~m from the interaction point.
It covers the pseudorapidity range $\eta > 7.9$.
The FNC is a lead--scintillator sandwich calorimeter. 
It consists of two longitudinal sections: the Preshower Calorimeter 
with a  length corresponding to about $60$ radiation lengths, 
or $1.6$ hadronic interaction lengths $\lambda$, 
and the Main Calorimeter with a total length 
of $8.9 \lambda$ (see figure~\ref{fig:FNC}a).
The acceptance of the FNC is defined by the aperture of the HERA
beam-line magnets and is limited to scattering angles of
$\theta\lsim 0.8$~mrad with approximately $30\%$ azimuthal coverage,
as illustrated in figure~\ref{fig:FNC}b.

The absolute electromagnetic and hadronic energy scales of the FNC are known to 
$5\%$ and $2\%$ precision, respectively~\cite{Aaron:2010ab}.
The energy resolution of the FNC calorimeter for electromagnetic
showers is $\sigma(E)/E \approx 20\%/\sqrt{E~[\rm GeV]} \oplus 2\%$
and for hadronic showers $\sigma(E)/E \approx 63\%/\sqrt{E~[\rm GeV]} \oplus 3\%$,
as determined in test beam measurements. 
The spatial resolution is
$\sigma(x,y)\approx 10\rm cm/\sqrt{E~[\rm GeV]} \oplus 0.6~\rm cm$ for 
hadronic showers starting in the Main Calorimeter.
A better spatial resolution of about $2\rm~ mm$ is achieved for
electromagnetic showers and for those hadronic showers 
which start in the Preshower Calorimeter.

The instantaneous luminosity is monitored based on the rate of the
Bethe-Heitler process $ep \to ep\gamma$. The final state photon is detected in the
photon detector located close to the beampipe at $z=-103$ m.
The precision of the integrated luminosity measurement is improved
in a dedicated analysis of the elastic QED Compton process~\cite{Lumi_QED}
in which both the scattered electron and the photon are detected in the SpaCal. 

\subsection{Event selection}
\label{sec:sample}
%
%
The data sample of this analysis has been collected using a special
low multiplicity 
trigger requiring two tracks
with $p_T>160$ MeV and originating from the nominal event vertex,
and at most one extra track with $p_T>100$ MeV.
The tracks are found by the Fast Track Trigger (FTT)~\cite{FTT}, 
based on hit information provided by the CJCs.
The trigger also contains a veto condition against non-$ep$ background
provided by the CIP.
The average trigger efficiency is about $75 \%$ for the analysis phase space.
The trigger simulation has been verified and tuned to the data 
using an independently triggered data sample.

%
For the analysis, exclusive events are selected, containing two oppositely 
charged pion candidates in the central tracker, a leading neutron in the FNC 
and nothing else above noise level in the detector\footnote{According
       to simulation, the forward going $\pi^+$ from the proton vertex 
       is emitted in the range $\eta > 5.7$ where it cannot be 
       reliably measured or identified with the available apparatus.}.
%
%
The photoproduction regime is ensured by the absence of a high energy 
electromagnetic cluster consistent with a signal from a scattered 
beam positron in the calorimeters.
This limits the photon virtuality to $\qsq \lesssim 2 \, \gevsq$, 
resulting in a mean value of $\langle \qsq \rangle = 0.04 \, \gevsq$.


The $\rho^0$ candidate selection requires the reconstruction 
of the trajectories of two, and only two, 
oppositely charged particles in the central tracking detector. 
They must originate from a common vertex 
lying within $\pm 30$ cm in $z$ of the nominal $ep$ interaction point, 
and must have transverse momenta above $0.2$ GeV and polar angles 
within the interval $ 20^{\circ} \leq \theta \leq 160^{\circ}$.
The momentum of the $\rho^0$ meson is calculated as the vector sum 
of the two charged particle momenta.
The two-pion invariant mass is required to be 
within the interval $0.3 < M_{\pi\pi}<1.5$ GeV.
Since no explicit hadron identification is used,
events are discarded with $M_{KK}<1.04$ GeV
where $M_{KK}$ is the invariant mass of two particles under the 
kaon mass hypothesis. This cut suppresses a possible background
from exclusive production of $\phi$ mesons.


Events containing a leading neutron are selected by requiring a
hadronic cluster in the FNC with an energy above $120~\GeV$ and a
polar angle below $0.75~\mathrm{mrad}$.
The cut on polar angle, defined by the geometrical acceptance of the FNC,
restricts the neutron transverse momenta to the range
$\ptn< x_L \cdot 0.69~\GeV$.

            
To ensure exclusivity, additional cuts are applied on the
calorimetric energy and on the response of the forward detectors.
There should be no cluster with energy above $400$ MeV, 
unless associated with $\rho^0$ decay products, 
in the SpaCal and LAr calorimeters.
A Large Rapidity Gap signature is required, by selecting events with 
no activity above noise levels in the forward detectors.
This suppresses non-diffractive interactions to a negligible level
and also significantly reduces diffractive background.  

\begin{table}[hbt]
        \centering
        \renewcommand{\arraystretch}{1.50}
\bt{|l r|c|c|}
 \hline 
     \multicolumn{2}{|l|}{Event selection $(2006\!-\!2007,~e^+p)$} &
                                             Analysis PS   &    Measurement PS           \\
 \hline 
    Trigger {\tt s14} (low multiplicity)& &                &                             \\
    No $e'$ in the detector    &  &   $Q^2 < 2$ GeV$^2$    &    $Q^2 = 0$ GeV$^2$        \\
 \hline 
    $2$ tracks, net charge $=0$, & &                       &                             \\
    ~ $p_T\!>\!0.2$ GeV,~ $20^o\!<\!\theta\!<\!160^o$,     &
                                &  $20<\Wgp<100$ GeV       &  $20<\Wgp<100$ GeV          \\
    ~ from $|z_{\rm vx}|<30$ cm & & $\ptr<1.0$ GeV         &  $-t^{\prime}<1.0$ GeV$^2$  \\
    $0.3<M_{\pi\pi}<1.5$ GeV    & & $0.6<M_{\pi\pi}<1.1$ GeV & $2m_{\pi}<M_{\rho}<
                                                            M_{\rho}\!+\!5\Gamma_{\rho}$ \\
 \hline 
         LRG requirement       & & $\sim 637,000$ events   &                             \\
 \hline 
    $E_n>120$ GeV              & & $x_L > 0.2$             & $0.35<x_L<0.95$             \\
    $\theta_n<0.75$ mrad       & & $\theta_n<0.75$ mrad    & $\ptn<x_L\cdot 0.69$ GeV    \\
 \hline 
    $\sim 7000$ events       &  & $\sim 6100$ events       &  $\sim 5770$ events         \\
 \hline 
 \hline 
  (OPE dominated range) & 
     OPE1 & \multicolumn{1}{l}{$\ptn\!<\!0.2$ GeV}     & \multicolumn{1}{r|}{($\sim 3600$ events)} \\
   & OPE2 & \multicolumn{2}{l|}{$\ptn\!<\!0.2$ GeV,~ $0.65\!<\!x_L\!<\!0.95$~~($\sim 2200$ events)}\\
 \hline 
\et
  \vspace{\baselineskip} 
        \caption{Event selection criteria and the definition 
                 of the kinematic phase space (PS) of the measurements.
                 The measured cross sections are determined at $Q^2=0$
                 using the effective flux~(\ref{eq:gflux}), based on the VDM.}
  \label{tab:selection}
\end{table}


After these cuts the data sample contains about $7000$ events.
The event selection criteria together with the analysis and the measurement 
phase space definitions are summarised in table~\ref{tab:selection}. 
In order to better control migration effects and backgrounds 
most of the selection cuts are kept softer
than the final measurement phase space limits.
In the end, the $\gp$ cross sections measured in the $\theta_n<0.75$ mrad 
range are based on $\sim\!5770$ events.
For the $\gpi$ cross section extraction additional cuts are applied
in order to stay within a range where the validity of OPE 
can be safely expected. Two sub-samples are defined:
 OPE1 with $\ptn<200$ MeV, containing $\sim\!3600$ events and
 OPE2 with $\ptn<200$ MeV and $x_L>0.65$, containing $\sim\!2200$ 
      events\footnote{The OPE2 sample corresponds 
      to the low $|t|<0.2$ GeV$^2$ region, see figure~\ref{fig:FNC}c.}.

\subsection{Monte Carlo simulations and corrections to the data}
\label{sec:mc}

Monte Carlo (MC) simulations are used to calculate acceptances and efficiencies 
for triggering, track reconstruction, event selection and background contributions
and to account for migrations between measurement bins due to the finite 
detector resolution.

 Signal events from the DPP reaction (figure~\ref{fig:FD}a)
 are modelled by the two-step MC generator POMPYT~\cite{Pompyt},
 in which the virtual pion is produced at the proton vertex 
 according to one of the available pion flux parametrisations.
 This pion then scatters elastically on the photon from the
 electron beam, thus producing a vector meson ($\rho^0$ in our case).
 In this analysis the non-Reggeized pion flux factor is taken 
 from the light-cone representation~\cite{light_cone} with the form factor 
 in equation~(\ref{eq:piflux})
%
\beq
   F^2(t,x_L) = \exp \left(- R^2_{\pi n} \frac{m^2_\pi-t}{1-x_L}\right),
  \label{eq:piff}
\eeq
%
 where $R_{\pi n}=0.93~\GeV^{-1}$ is the radius of the pion-proton 
 Fock state~\cite{flux_Holt}.
 The same version of the pion flux factor has been used in previous H1 publications
 on leading neutron measurements~\cite{Aktas:2004,Aaron:2010ab}
 providing a good description of inclusive neutron spectra.
 For the numerical value of the $p\pi n$ coupling constant, the most recent
 estimate~\cite{gpiPN}  $g_{p\pi n}^2/4\pi = 14.11 \pm 0.20$ is used.  

 Since the exact shape of the $\ptr^2$ dependence is not {\em a priori} known,
 two extreme versions are generated. 
 In the first version a simple exponential shape is assumed, as
 expected for elastic $\rho^0$  photoproduction on the pion,
 with the slope $b=5$ GeV$^{-2}$.
 For the second version a mass-dependent slope is taken, 
 $4\leq b(M_{n\pi})\leq 22$ GeV$^{-2}$,
 typical for DPP processes as observed at lower energies~\cite{ISR1,ISR2,Zotov}.
 The difference in the correction factors obtained using these two versions of 
 MC simulations is part of the model dependent systematic uncertainty.

The background events originating from diffractive $\rho^0$ production
(figure~\ref{fig:FD}d) are generated using the program DIFFVM~\cite{DiffVM}, 
which is based on Regge theory and the Vector Dominance Model. 
All channels (elastic, single- and double-dissociation processes) are included,
with the relative composition as measured in~\cite{H1_stotgp}.
For the proton dissociative case the $M_Y$ mass spectrum is parametrised as
$\mathrm{d} \sigma / \mathrm{d}M_Y^2 \propto 1/M_Y^{2.16}$,
for $M_Y^2 > 3.6~\gevsq$ with quark and diquark fragmentation using the JETSET
program~\cite{jetset}.
For the low mass dissociation the production of excited nucleon states 
at the proton vertex is taken into account explicitly. Signal
events, corresponding to the diagram shown in figure~\ref{fig:FD}c, are excluded
from the generated background sample.

The DIFFVM program is also used to estimate possible contaminations
from diffractive $\omega(782),~ \phi(1020)$ and $\rho^{\prime}(1450-1700)$
production.

As discussed in section~\ref{sec:intro}, the pion exchange diagram dominates
the cross section in the low $t$ region where the contributions from the diagrams 
in figures~\ref{fig:FD}b and \ref{fig:FD}c almost cancel. 
To check a possible influence of these terms on the MC correction factors,
neutron exchange events (b) were generated using POMPYT and events
of class (c) using DIFFVM. As expected, these events have
kinematic distributions and selection efficiencies similar to those 
from the pion exchange process and do not alter the MC correction factors
beyond the quoted systematic uncertainties.
   


In both the POMPYT and the DIFFVM generators a simple non-relativistic 
Breit-Wigner shape is used for the $\rho$ meson mass. Therefore all MC events 
are reweighted to the relativistic Breit-Wigner shape with additional 
$p_T$-dependent distortion as observed in $\rho^0$ photoproduction experiments.
The distortion is caused by the interference between the resonant 
and non-resonant $\pi^+\pi^-$ production 
and is characterised by the phenomenological skewing parameter,
$n_{RS}$, as suggested by Ross and Stodolsky~\cite{RS}:
%
\beq
     \frac{dN(M_{\pi\pi})}{dM_{\pi\pi}} \propto BW_{\rho}(M_{\pi\pi})
     \left( \frac{M_{\rho}}{M_{\pi\pi}}\right)^{n_{RS}(\ptr)}
   \label{eq:RS}
\eeq
%
with $M_{\rho}$ being the nominal resonance mass~\cite{pdg} 
and $n_{RS}(\ptr)$ taken from published ZEUS data
on elastic photoproduction of $\rho^0$ mesons~\cite{ZEUS_rho}.   
Additionally, the signal MC events (POMPYT) are reweighted in
$\Wgp$ and in $\ptr^2$ to the observed shapes of the 
corresponding distributions. This reweighting is performed
iteratively and has converged after two iterations.
The uncertainty in the reweighting procedure
is then taken into account in the systematic error analysis.

Small, but non-zero values of $Q^2$ cause $|t'|$ to differ 
from $\ptr^2$ by less than $Q^2$. 
To account for this effect 
a multiplicative correction factor determined 
with the Monte Carlo generators is applied to the bins of 
the $\ptr^2$ distribution;
the correction is obtained by taking the ratio between the
$|t'|$ and $\ptr^2$ distributions at the generator level.
This correction varies from 1.1 at $\ptr^2 = 0$ to 0.77 at
$\ptr^2 = 1$~GeV$^2$.

For all MC samples detector effects are simulated in detail with the 
GEANT program~\cite{Geant3}. 
The MC description of the detector response, including trigger efficiencies, 
is adjusted using comparisons with independent data. 
Beam-induced backgrounds are taken into account by overlaying
the simulated events with randomly triggered real events.
The simulated MC events are passed through the same 
reconstruction and analysis chain as is used for the data.

The MC simulations are used to correct the distributions 
at the level of reconstructed particles
back to the hadron level on a bin-by-bin basis.
The size of the correction factors is $12$ in average,
corresponding to an efficiency of $\sim\!8\%$, and varies 
between $\sim\!10$ and $\sim\!24$ for different parts of the
covered phase space.
The main contributions to the inefficiency are: the azimuthal
acceptance of the FNC ($\sim\!30\%$ on average), the $\rho$ meson
reconstruction efficiency which is zero if one of the tracks has low
transverse momentum ($\sim\!60\%$), the LRG selection efficiency ($\sim\!60\%$) and the
trigger efficiency ($\sim\!75\%$).
The bin purity, defined as the fraction
of events reconstructed in a particular bin that originate from
the same bin on hadron level,  varies between $70\%$ and $95\%$
for one-dimensional distributions and between $45\%$ and $65\%$ 
for two-dimensional ones.
As an example, figure~\ref{fig:FNC}c illustrates the binning scheme
used in the two-dimensional $(x_L,\ptn)$ distribution. 

\subsection{Extraction of the $\rho^0$ signal}
\label{sec:rho}
\begin{figure}[ppp]
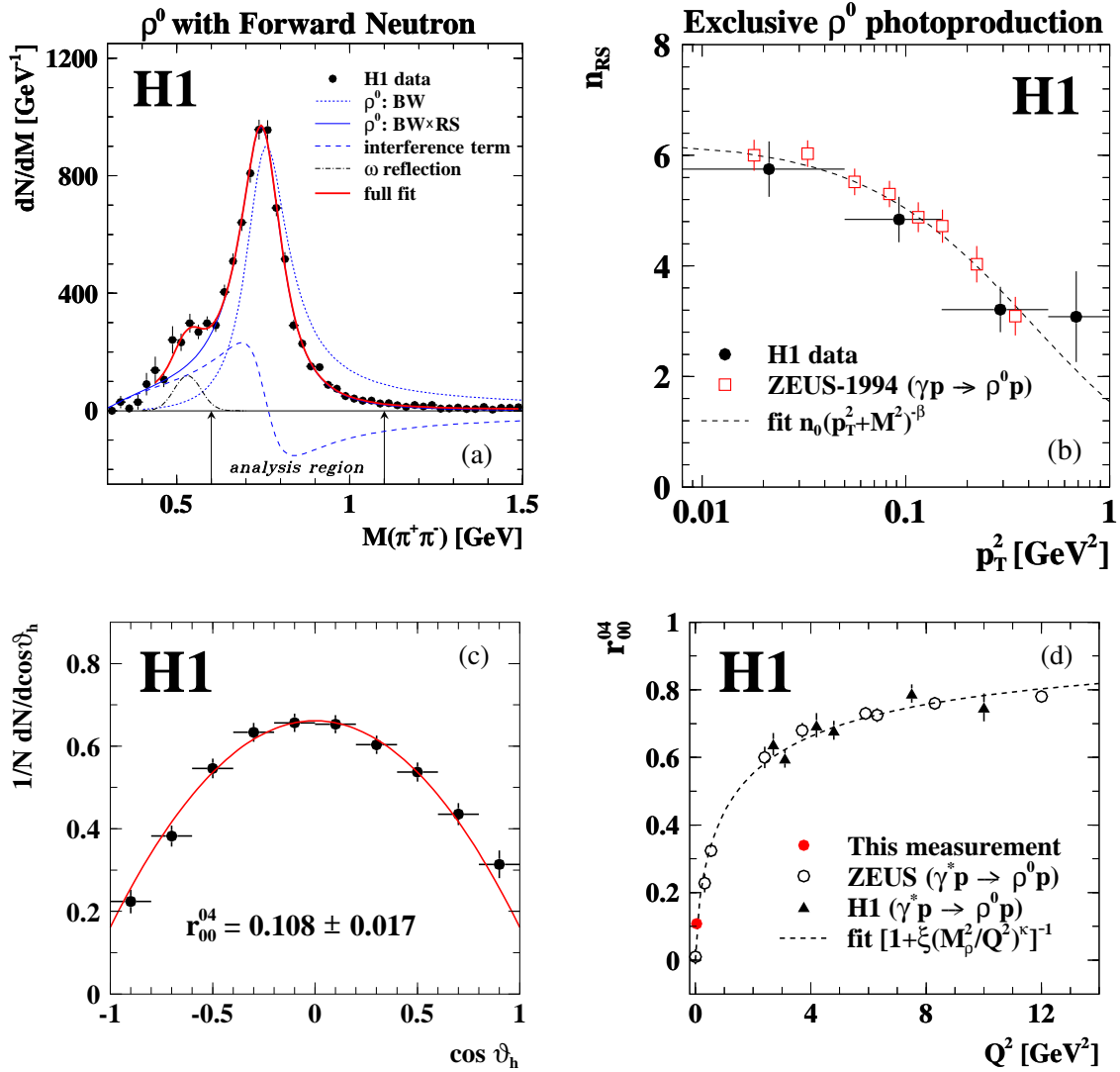

\center
 \epsfig{file=d15-120f3a.eps,width=0.47\textwidth}
 \epsfig{file=d15-120f3b.eps,width=0.49\textwidth}
 \epsfig{file=d15-120f3c.eps,width=0.48\textwidth}
 \epsfig{file=d15-120f3d.eps,width=0.48\textwidth}
 \setlength{\unitlength}{1cm}
\caption{The $\rho^0$ meson properties:
(a) Mass distribution of the $\pi^+\pi^-$ system for exclusive
         $\rho^0$ production with $p_T^2<1.0$ GeV$^2$ associated
         with a leading neutron.
         The data points are corrected for the detector efficiency.
         The curves represent different components contributing
         to the measured distribution and the Breit-Wigner resonant part
         extracted from the fit to the data.
         The analysis region $0.6<M_{\pi^+\pi^-}<1.1$ GeV is indicated
         by vertical arrows. 
         (b) Ross-Stodolsky skewing parameter, $n_{RS}$, as a function
         of $p_T^2$ of the $\pi^+\pi^-$ system. 
         The values measured in this analysis
         are compared to previously obtained results 
         for elastic photoproduction of $\rho^0$ mesons, 
         $\gamma p \to \rho^0 p$, by the ZEUS Collaboration.
         (c) Decay angular distribution of the $\pi^+$ in the helicity frame.
         (d) Spin-density matrix element, $r_{00}^{04}$, as a function
         of $Q^2$ for diffractive $\rho^0$ photo- and
         electro-production.
         The curves on figures (b-d) represent the results of the fits
         discussed in the text.}
\label{fig:mass}
\end{figure}

The invariant mass distribution of the two tracks under 
the charged pion mass hypothesis is shown in figure~\ref{fig:mass}a.
The distribution is corrected for the mass dependent detector
efficiency. 

A fit is performed in the range $M_{\pi\pi}>0.4$ GeV
using the Ross-Stodolsky parametrisation~(\ref{eq:RS})
for the $\rho^0$ meson mass shape and adding the contributions 
for the reflection from $\omega \rightarrow \pi^+\pi^-\pi^0$
and for the non-resonant background. 
Other sources of non-$\rho^0$ background, such as 
$\omega(782)   \to  \pi^+\pi^-, ~~ 
 \phi(1020)    \to  K_L^0 K_S^0, \, \pi^+\pi^-\pi^0, ~~
 \rho^{\prime} \to  \rho\pi\pi, 4\pi, \pi\pi$,    
which may be misidentified as $\rho^0$ candidates,
are estimated using MC simulations with the relative yield
normalisation fixed to previously measured and published values:
$\sgp(\omega)/\sgp(\rho^0)=0.10(\pm20\%)$~\cite{ZEUS_omega}, 
$\sgp(\phi)/\sgp(\rho^0)=0.07(\pm20\%)$~\cite{ZEUS_phi} and 
$\sgp(\rho^{\prime})/\sgp(\rho^0)=0.20(\pm50\%)$~\cite{rho_prime}. 
The resulting  overall background contamination in the
analysis region $0.6<M_{\pi\pi}<1.1$ GeV is found to be
$(1.5 \pm 0.7)\%$.  

The fitted values of the resonance mass and width are 
$764 \pm 3 (\rm stat.)$ MeV and $155 \pm 5 (\rm stat.)$ MeV, respectively, 
in agreement with the nominal PDG values of $M_{\rho}$ and $\Gamma_{\rho}$~\cite{pdg}.
The cross section is then calculated for the full mass range
$2m_{\pi}<M_{\pi\pi}<M_{\rho}+5\Gamma_{\rho}$ using the resonant part only, 
represented by the relativistic Breit-Wigner function $BW_{\rho}(M_{\pi\pi})$
with momentum dependent width $\Gamma(M_{\pi\pi})$~\cite{BW_rel}:
%
\beq
     BW_{\rho}(M_{\pi\pi}) = 
     \frac{M_{\pi\pi} \, M_{\rho} \, \Gamma(M_{\pi\pi})}
          {(M_{\rho}^2-M_{\pi\pi}^2)^2 + M_{\rho}^2 \, \Gamma(M_{\pi\pi})^2}, 
     ~~~~~
     \Gamma(M_{\pi\pi}) = 
     \Gamma_{\rho} \, \left( \frac{q^*}{q_0^*} \right)^3 \,
        \frac{M_{\rho}}{M_{\pi\pi}}
   \label{eq:BWR}
\eeq
%
where 
$q^*$ is the momentum of the decay pions in the rest frame of a pair of pions 
with mass $M_{\pi\pi}$, and $q^*_0$ is the value of $q^*$ 
for $M_{\pi\pi}=M_{\rho}$.

The Breit-Wigner shape is strongly distorted due to interference 
with the non-resonant $\pi\pi$ production amplitude 
(dashed curve in figure~\ref{fig:mass}a). 
The strength of the distortion is $p_T$-dependent
and within the ansatz~(\ref{eq:RS}) is
characterised by the phenomenological skewing parameter, $n_{RS}$.
For the full $p_T$ range of the present analysis, $\ptr^2<1$ GeV$^2$,
a fit results in the value $n_{RS} = 4.22 \pm 0.28$.
To study its $p_T$ dependence the fit is repeated in four $\ptr^2$ bins.
The values obtained are shown in figure~\ref{fig:mass}b in comparison 
with previously published ZEUS results~\cite{ZEUS_rho}
from elastic $\rho^0$ photoproduction, $\gamma p \to \rho^0p$.
The dashed curve represents a fit to all these data
by the empirical formula
%
\beq
    n_{RS} = n_0 \, (p_T^2 + M^2)^{-\beta}
  \label{eq:nRS}
\eeq
%
with $n_0, M$ and $\beta$ as free parameters.
The fitted value of $M^2 \simeq 0.6$ GeV$^2$ suggests that 
the relevant scale for  photoproduction of vector mesons  
is indeed $(p_T^2 + M_V^2)$. 
 

An important set of observables which characterise the helicity structure 
of the vector meson production are the angular distributions of the decay pions.
Here we study the distribution of $\theta_h$ which gives access to the
$\rho^0$ spin-density matrix element $r_{00}^{04}$.
The angle $\theta_h$ is defined as the polar angle of the positively charged 
decay pion in the $\rho^0$ rest frame with respect to the meson
direction in the $\gsp$ centre-of-mass frame.
According to the formalism presented in~\cite{helicity} the distribution
$\theta_h$ is given by:
%
\beq
    \frac{1}{\sigma}\frac{{\rm d}\sigma}{{\rm d}\cos \theta_h} 
    \propto 1-r_{00}^{04} +(3 r_{00}^{04}-1) \cos^2 \theta_h .
  \label{eq:decay}
\eeq
%
Figure~\ref{fig:mass}c shows the acceptance corrected $\cos \theta_h$
distribution together with the fit by equation~(\ref{eq:decay}) 
yielding the value of $r_{00}^{04} = 0.108 \pm 0.017$. 
In figure~\ref{fig:mass}d this result
is compared to the values obtained in diffractive $\rho^0$ photo- and
electro-production at HERA~\cite{ZEUS_rho,r0004_ZEUS,r0004_H1}.
The steep $Q^2$ dependence is driven mainly by the QED gauge invariance 
motivated factor $Q^2/M_{V}^2$, 
and can be fitted by a simple expression~\cite{INS_model}
%
$$
   r_{00}^{04} = \frac{1}{1 + \xi (M_{\rho}^2/Q^2)^{\kappa}}
$$
%
with the parameters $\xi = 1.85 \pm 0.10$ and $\kappa = 0.67 \pm 0.03$,
as illustrated by the dashed curve.

In summary, all properties of the selected $\pi^+\pi^-$ sample
investigated here are consistent with $\rho^0$ photoproduction.

\subsection{Signal and background decomposition}
\label{sec:s2b}
The event selection described in section~\ref{sec:sample}
does not completely suppress non-DPP background. 
According to the MC simulations, the remaining part
is mostly due to proton dissociation with some admixture of double
dissociative events. 

\begin{figure}[ppp]
\center
 \epsfig{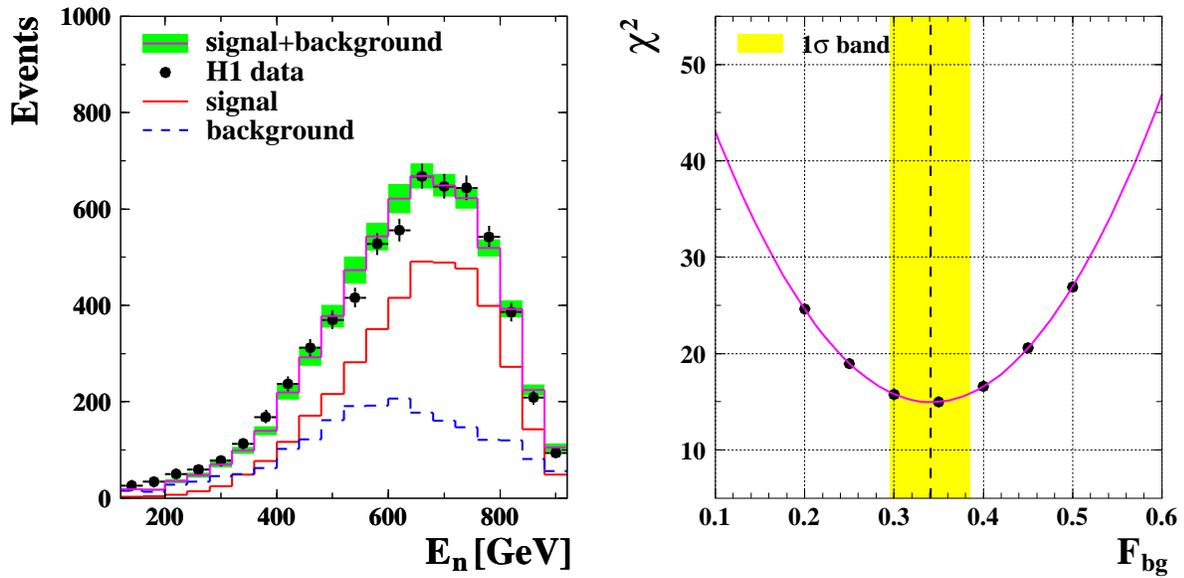}
 \setlength{\unitlength}{1cm}
\caption{Signal to background decomposition in the selected data sample.
         On the left panel the distribution of the measured neutron energy, $E_n$,
         is shown together with the contributions from signal and background.
         On the right panel the $\chi^2$ dependence on the background fraction,
         $F_{bg}$, is shown. 
         The shaded band represents the $1 \sigma$ uncertainty 
         around the optimal fit value of the $F_{bg}$, taking into account
         statistical errors, FNC calibration systematics and the uncertainty
         in proton dissociation background shape.}
\label{fig:S2B}
\end{figure}

As in the case of inclusive leading neutron production~\cite{Aaron:2010ab} 
signal and background events
have different shapes of the leading neutron energy distribution,
although in the present analysis the difference is less pronounced.
%
%
The shape differences in the neutron energy spectrum predicted by MC 
for the DPP events (POMPYT) and for the proton dissociative
background (DIFFVM) are still sufficient to disentangle 
these two contributions on a statistical basis. 
For this purpose a combination of the spectra obtained for 
reconstructed events of these two MC models fulfilling all selection criteria   
with free normalisation is fitted to the data. From this fit the background 
fraction is determined to be $F_{bg}=0.34\pm 0.05$.
The uncertainty includes both the fit error and systematic uncertainties
related to the background shape variation in terms of $M_Y$ and $t$ 
dependencies and proton dissociation fraction in the overall 
diffractive cross section.
Figure~\ref{fig:S2B} illustrates this decomposition using the nominal
DIFFVM parameters.
%

\begin{figure}[ppp]
\center
 \epsfig{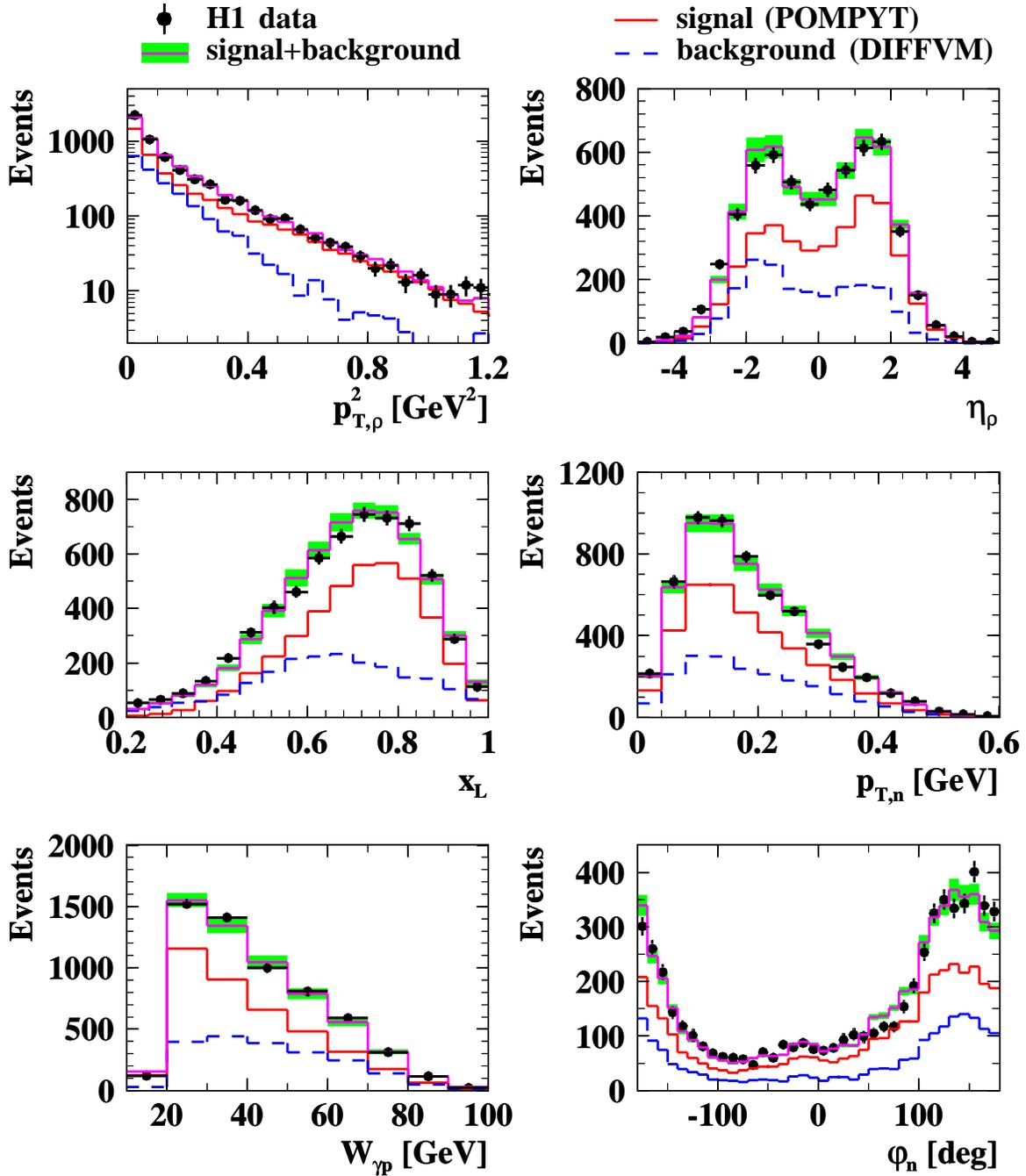}
 \setlength{\unitlength}{1cm}
\caption{Distributions of the reconstructed quantities $p_T^2$ and 
         $\eta$ of the $\rho^0$ meson, $x_L$, $p_T$ and $\varphi$ 
         of the neutron and $\Wgp$ for data
         and Monte Carlo simulations normalised to the data.
         Data points are shown with statistical errors only.
         The shaded band indicates the uncertainty 
         in the estimated background fraction.}
\label{fig:CP}
\end{figure}

Control plots for the data description by the Monte Carlo models 
using this signal to background ratio are shown in figure~\ref{fig:CP}.
Since neither POMPYT nor DIFFVM are able to provide reliable 
absolute cross section predictions for such a final state, 
only a shape comparison is possible.
The irregular shape of the azimuthal angle distribution, $\varphi_n$, is
due to the FNC aperture limitations, as shown in figure~\ref{fig:FNC}b.

In the fit described above the absolute normalisation for the DIFFVM 
prediction is left free. As a cross check, this normalisation
has been fixed using an orthogonal, background dominated sample,
obtained by requiring an `anti-LRG' selection, i.e. $\rho^0 + n$ 
events with additional activity in the forward detectors.
In this sample the background fraction is found to be $0.58\pm 0.07$.
Fixing the DIFFVM normalisation 
by a fit to the `anti-LRG' sample results in a background contribution
of $F_{bg}=0.29\pm 0.05$ in the main sample.
Since the signal-to-background decomposition fit 
      in this cross check gives a worse 
      $\chi^2$, the nominal value $F_{bg}=0.34$
      is used for the cross section determinations.
The difference to the $F_{bg}$ value determined in the nominal analysis,
as described above, is well covered by systematic uncertainty 
of the LRG condition efficiency.

\subsection{Cross section determination and systematic uncertainties}
\label{sec:sys}

The cross sections are measured for the kinematic ranges as
defined in the rightmost column of table~\ref{tab:selection}.
From the observed number of $ep$ events, $N_{data}$, the bin-integrated
$\gamma p$ cross section in bin $i$ is calculated as
%
\beq
     \sgp^i = \frac{1}{\Phi_{\gamma}} 
              \frac{N_{data}^i - N_{bg}^i}{{\cal L}(A \cdot \epsilon)_i}
              \cdot C_{\rho}^i
  \label{eq:xsec}
\eeq
%
where $N_{bg}^i$ is the expected diffractive dissociation background in bin $i$
taking into account the overall normalisation fraction $F_{bg}=0.34$,
$A \cdot \epsilon$ is the correction for detector acceptance and efficiency,
${\cal L}$ is the integrated luminosity of the data, 
$C_{\rho}$ is the extrapolation factor for the number of $\rho^0$ events
from the $M_{\pi\pi}$ measurement interval to the full $\rho^0$ mass range
and $\Phi_{\gamma}=0.1543$ is the value of the equivalent photon flux
from equation~(\ref{eq:gflux})
for the given $(\Wgp,Q^2)$ range\footnote{ Note, 
   that the effective VDM flux~(\ref{eq:gflux}) converts 
   the $ep$ cross section into a {\it real} $\gamma p$ cross section 
   at $Q^2=0$, contrary to the EPA flux~\cite{EPA} converting it to the 
   transverse $\gamma^*p$ cross section, averaged over the measured 
   $Q^2$ range. The difference between the two approaches amounts to 
   $\approx 6\%$ integrated over the $(Q^2,y)$ range of the measurement.}.
Since the statistics available does not allow for a reliable $\rho^0$ 
mass fit in every measurement bin, $C_{\rho}^i$ is calculated
using $C_{\rho}=1.155$, obtained from the fit of the full sample
and bin-dependent skewing correction factor derived from
the fitted dependence of $n_{RS}(\ptr^2)$ in equation~(\ref{eq:nRS}). 
 
Several sources of experimental uncertainties are considered and
their effects on the measured cross section are quantified. The systematic
uncertainties on the cross section measurements are determined using
MC simulations, by propagating the corresponding
uncertainty through the full analysis chain.
The individual systematic uncertainties are grouped into four categories below.
\bi
  \item {\bf Detector related sources.} \\
         The trigger efficiency is verified and tuned  
         with the precision of $3.4\%$ using
         an independent monitoring sample. 
         It is treated as correlated between different bins. 

         The uncertainty due to the track finding and reconstruction 
         efficiency in the central tracker is estimated to be 
         $1\%$ per track~\cite{track_eff}
         resulting in $2\%$ uncertainty in the cross section,
         taken to be correlated between bins.

         Several sources of uncertainties related to the measurement
         of the forward neutrons are considered.
         The uncertainty in the neutron detection efficiency 
         which affects the measurement in a global way
         is $2\%$~\cite{Aaron:2010ab}.
         The $2\%$ uncertainty on the absolute hadronic energy scale 
         of the FNC~\cite{Aaron:2010ab} leads to a systematic error
         of $1.1\%$ for the $x_L$-integrated cross section and 
         varying between $2\%$ and $19\%$ in different $x_L$ bins. 
         The acceptance of the FNC calorimeter is defined 
         by the interaction point and the geometry of the HERA magnets
         and is determined using MC simulations.
         The uncertainty of the impact position of the particle on the FNC,
         due to beam inclination and the uncertainty on the FNC position,
         is estimated to be $5$~mm \cite{Aaron:2010ab}.
         This results in an average uncertainty 
         on the FNC acceptance determination of $4.5\%$ reaching up to $10\%$
         for the $\ptn$ distribution. 

         The systematics due to the exclusivity condition in the main 
         part of the H1 detector is estimated to be $2.1\%$. It gets
         contributions from varying the LAr calorimeter noise cut between
         $400$ MeV and $800$ MeV ($0.9\%$) and from the parameters
         of the algorithm connecting clusters with tracks ($1.9\%$).
         This error influences only the overall normalisation.

         The uncertainty from the LRG condition is determined in the same 
         manner as in the H1 inclusive diffraction analyses based on
         the large rapidity gap technique~\cite{H1_F2D, H1_FLD}. It is
         further verified by comparing the cross sections obtained 
         using different components of the forward detector apparatus
         for the LRG selection:
         FMD alone vs FMD$+$FTS vs FMD$+$Plug vs FMD$+$FTS$+$Plug.
         The resulting uncertainty is conservatively estimated to be $9.0\%$
         affecting all bins in a correlated manner. 
  
\item {\bf Backgrounds.} \\ 
         Three different types of background are considered.

         Non-$ep$ background is estimated from the shape of 
         the $z$-vertex distribution and from the analysis 
         of non-colliding proton bunches to be $(1.2 \pm 0.7)\%$.
         Background originating from random coincidences between
         $\rho^0$ photoproduction events and neutrons from $p$-gas
         interactions amounts to $(1.0 \pm 0.2)\%$. This results in
         $2.2\%$ background which was statistically subtracted in
         all distributions with an uncertainty of $0.8\%$.
 
         Non-$\rho^0$ background, as discussed in section~\ref{sec:rho},
         has an uncertainty of $0.7\%$ and affects the overall 
         normalisation only.

         Diffractive background to the DPP signal events (section~\ref{sec:s2b})
         is estimated  with a precision of $7.6\%$. This is one 
         of the largest individual uncertainties in the analysis.
         It is correlated between the bins.

\item {\bf MC model uncertainties.} \\
         The uncertainty in the subtracted diffractive background
         due to the limited knowledge on $\gamma p$ diffraction 
         is evaluated by varying the $M_Y$ and $t$ dependencies 
         in the DIFFVM simulation and the relative composition
         of diffractive channels within the limits allowed by 
         previous HERA measurements. The resulting uncertainty is a part
         of the background subtraction systematics listed above.

         The systematic uncertainty of the MC correction factors for signal events
         is $4.1\%$, varying between $1\%$ and $9\%$ in different bins.
         It is evaluated from the difference between two versions
         of the POMPYT MC program with different $p_T^2$ dependencies
         of the $\rho^0$ cross section, as described in section~\ref{sec:mc}.
         Here the uncertainty due to the POMPYT reweighting procedure 
         is also accounted for.

\item {\bf Normalisation uncertainties.} \\
         The uncertainty related to the $\rho^0$ mass fit,
         extrapolating from the measurement domain
         $0.6 \leq M_{\pi\pi} \leq 1.1~\gev$ to the full mass range
         $2m_{\pi}<M_{\pi\pi}<M_{\rho}+5\Gamma_{\rho}$,
         which implies a correction factor of $ C_{\rho}=1.155$ 
         on average in  equation~(\ref{eq:xsec}) 
         with an uncertainty of $1.6\%$ due to fit errors.

         The integrated luminosity of the data sample
         is known with $2.7\%$ precision~\cite{Lumi_QED}.

         Together with other normalisation errors listed above 
         the resulting total normalisation uncertainty amounts to $4.4\%$.
\ei

The systematic uncertainties shown in the figures and tables 
are calculated using the quadratic sum of all contributions, 
which may vary from point to point. 
They are larger than the statistical uncertainties 
in most of the measurement bins.

The total systematic uncertainty for the integrated $\gamma p$ cross section
is $14.6\%$ including the global normalisation errors.

\section{Results}

Total, single- and double-differential photoproduction cross sections
for the reaction $\gprho$ are measured in the kinematic range 
defined in table~\ref{tab:selection}. The photon-pion cross section,
$\sgpi = \sigma(\gamma\pi^+ \to \rho^0\pi^+)$, is extracted from
the differential cross section ${\rm d}\sgp/{\rm d}x_L$ 
using the pion flux~\cite{flux_Holt} integrated 
over the range $\ptn<0.2$ GeV. 
The results are summarised in tables~\ref{tab:table2}-\ref{tab:table9}
and are shown in figures~\ref{fig:dsdxl}-\ref{fig:sigma_gpi}. 

\subsection{$\gp$ cross sections}
\label{sec:res_gp}
\begin{figure}[ppp]
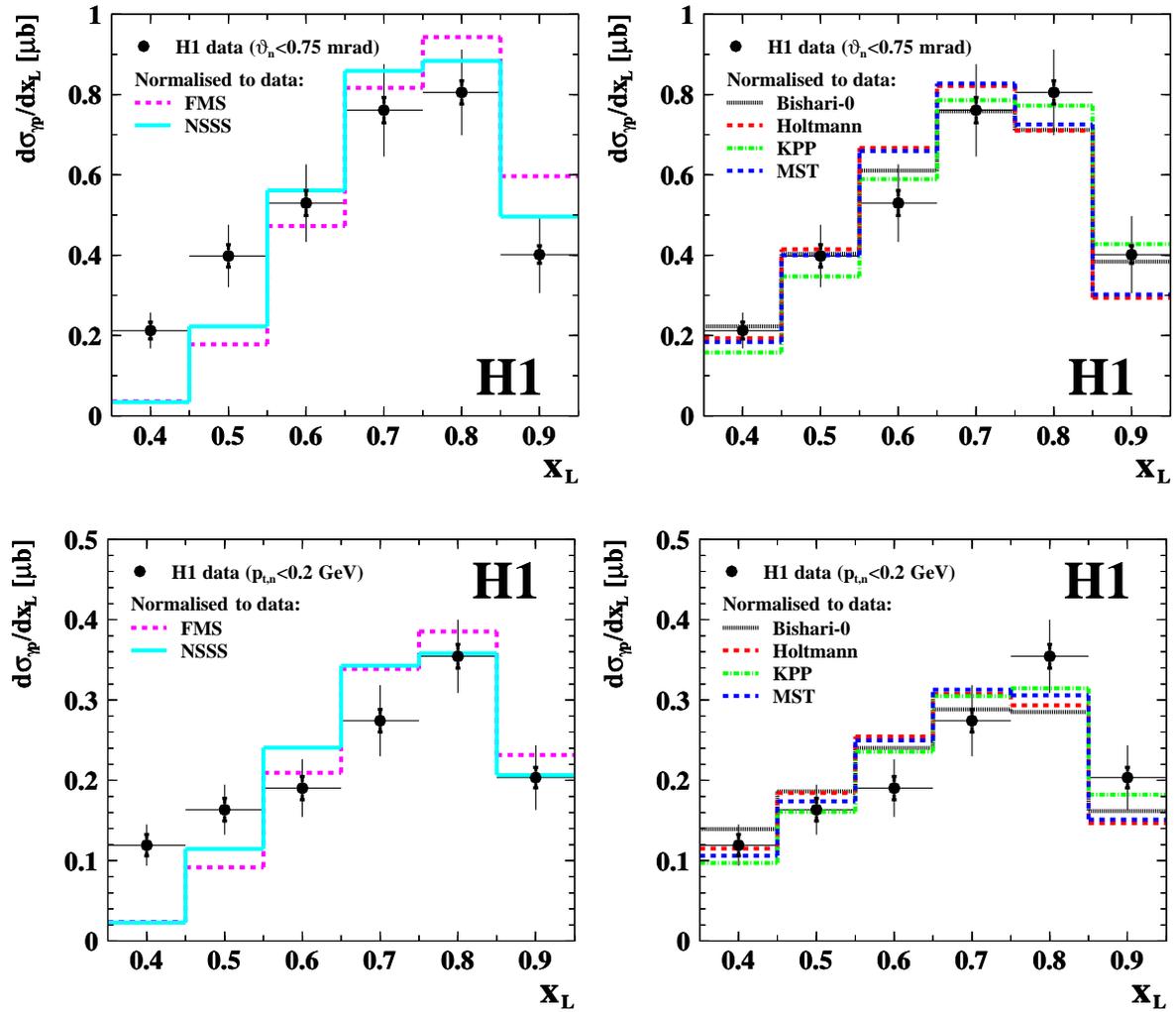

\center
 \epsfig{file=d15-120f6title.eps,width=\textwidth}
 \epsfig{file=d15-120f6a.eps,width=0.49\textwidth}
 \epsfig{file=d15-120f6b.eps,width=0.49\textwidth}
 \epsfig{file=d15-120f6c.eps,width=0.49\textwidth}
 \epsfig{file=d15-120f6d.eps,width=0.49\textwidth}
 \setlength{\unitlength}{1cm}
\caption{Differential cross section d$\sgp/{\rm d}x_L$ 
         in the range $20<\Wgp<100$ GeV compared
         to the predictions based on different versions of the pion flux models.
         Top row: cross sections in the full FNC acceptance $\theta_n<0.75$~mrad.
         Bottom row: cross sections for the OPE1 range, $\ptn<200$ MeV.
         Left-hand column: disfavoured versions of the pion fluxes, 
         right-hand column: pion fluxes compatible with the data. 
         The data points are shown with statistical (inner error bars) and total
         (outer error bars) uncertainties, excluding an overall normalisation
         error of $4.4\%$. All predictions are normalised to the data.}
\label{fig:dsdxl}
\end{figure}

The $\gp$ cross section integrated in the domain  
$0.35\!<\! x_L\!<\! 0.95$ and $-t^{\prime}\!<\! 1$ GeV$^2$
and averaged over the energy range $20 \!<\! \Wgp \!<\! 100$ GeV
is determined for two intervals 
of leading neutron transverse momentum:
\beq
  \sigma (\gamma p \to \rho^0 n \pi^+ ) = (310 \pm 6_{\rm stat} \pm 45_{\rm sys})~ {\rm nb}
         \hspace*{0.8cm} {\rm for} \hspace*{0.3cm} \ptn<x_L \cdot 0.69 {\rm ~GeV}
  \label{eq:sgp1}
\eeq
and 
\beq 
  \sigma (\gamma p \to \rho^0 n \pi^+ ) = (130 \pm 3_{\rm stat} \pm 19_{\rm sys})~ {\rm nb} 
         \hspace*{0.8cm} {\rm for} \hspace*{1.2cm} \ptn<0.2 {\rm ~GeV}.  
  \label{eq:sgp2}
\eeq
%
%
Single differential cross sections as a function of $x_L$ 
for these two regions are given in table~\ref{tab:table2} 
and are shown in figure~\ref{fig:dsdxl}. 
The data are compared in shape to the predictions based on different 
models for the pion flux. 
Some models, like FMS~\cite{flux_FMS} and NSSS~\cite{flux_NSSS} are disfavoured 
by the data and can be ruled out even on the basis of shape comparison alone.
The other pion flux parametrisations:
Bishari-$0$~\cite{flux_Bish}, Holtmann~\cite{flux_Holt},
KPP~\cite{flux_KPP} and MST~\cite{flux_MST}
are in good agreement with the data in both $\ptn$ ranges.

\begin{figure}[hhh]
\center
 \epsfig{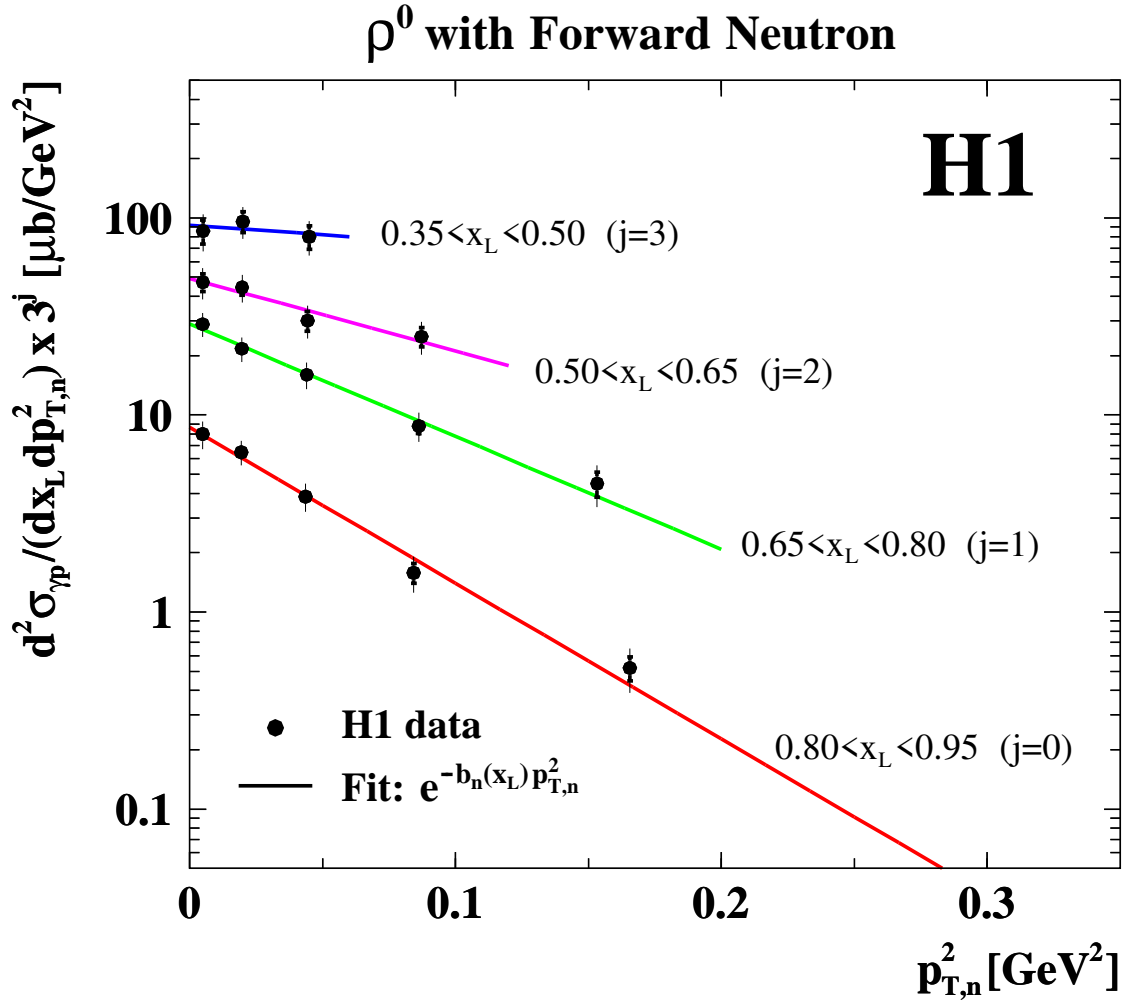}
 \setlength{\unitlength}{1cm}
\caption{Double differential cross section $\d2sxp$ of neutrons
         in the range $20<\Wgp<100$ GeV 
         fitted with single exponential functions.
         The cross sections in different $x_L$ bins $j$ are scaled by
         the factor $3^j$ for better visibility.
         The binning scheme is shown in figure~\ref{fig:FNC}c.
         The data points are shown with statistical (inner error bars)
         and total (outer error bars) uncertainties excluding 
         an overall normalisation error of $4.4\%$.
}
\label{fig:ddn}
\end{figure}

\begin{figure}[hhh]
\center
 \epsfig{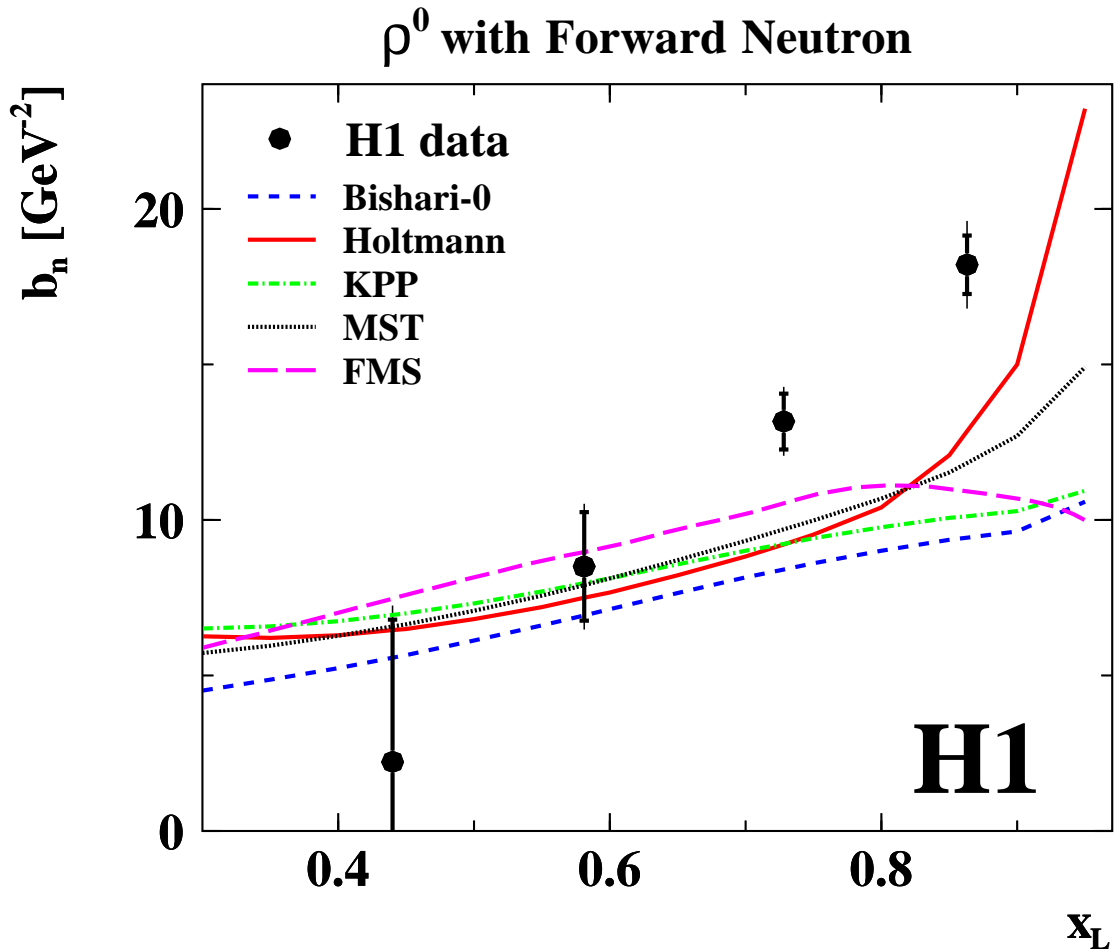}
 \setlength{\unitlength}{1cm}
\caption{The exponential slopes 
         fitted through the $p_T^2$ dependence of the leading neutrons
         as a function of $x_L$.
         The inner error bars represent statistical errors and the outer
         error bars are statistical and systematic errors added in quadrature.
         The data points are compared to the expectations of  
         several parametrisations of the pion flux within the
         OPE model.
}
\label{fig:bn_xl}
\end{figure}

Additional constraints on the pion flux models could be provided by
the dependence on $t$ (or $\ptn^2$) of the leading neutron. 
The double differential cross section $\d2sxp$ is measured, 
and the results are presented in table~\ref{tab:table3} and figure~\ref{fig:ddn}.
The bins are chosen such, that the data are not affected by the
polar angle cut (see figure~\ref{fig:FNC}c). 
Although neither the $t$-, nor the $\ptn^2$-dependence of the pion flux
models are exactly exponential 
they can be approximated by a simple exponent in many cases. 
Such an approximation has been used already
in other analyses~\cite{Chekanov:2002,Chekanov:2007}.
The $\ptn^2$-distributions measured here for fixed $x_L$ are compatible
with an exponential shape within the statistical 
and uncorrelated systematic errors. 
Therefore, the same approach is used here.
%
The cross sections are fitted by a single exponential function
$e^{-b_n(x_L)\ptn^2}$ in each $x_L$ bin. 
The quality of the fits is good, with $P(\chi^2)=0.35 \div 0.60$. 
The results are presented in table~\ref{tab:table4} 
and figure~\ref{fig:bn_xl}. The measured $b$-slopes are compared to
those obtained from several pion flux parametrisations.
Despite of the large experimental uncertainties none of the models 
is able to reproduce the data\footnote{Reweighting the signal MC using
  the measured $b_n(x_L)$ slopes 
  has only small effects on the cross section determination and is
  covered by the systematic uncertainties assigned to the pion flux
  models}.
A possible reason for this discrepancy could be the effect
of energy-momentum conservation affecting the proton vertex
in this exclusive reaction more strongly than in inclusive
production of a leading neutron in which an apparent factorisation
of the proton vertex has been observed.
Another explanation~\cite{Zotov,abs_corr_3} 
could be absorptive corrections
which modify the $t$ dependence of the amplitude, leading to an
increase of the effective $b$-slope at large $x_L$
as compared to the pure OPE model without absorption.

\begin{figure}[ppp]
\center
 \epsfig{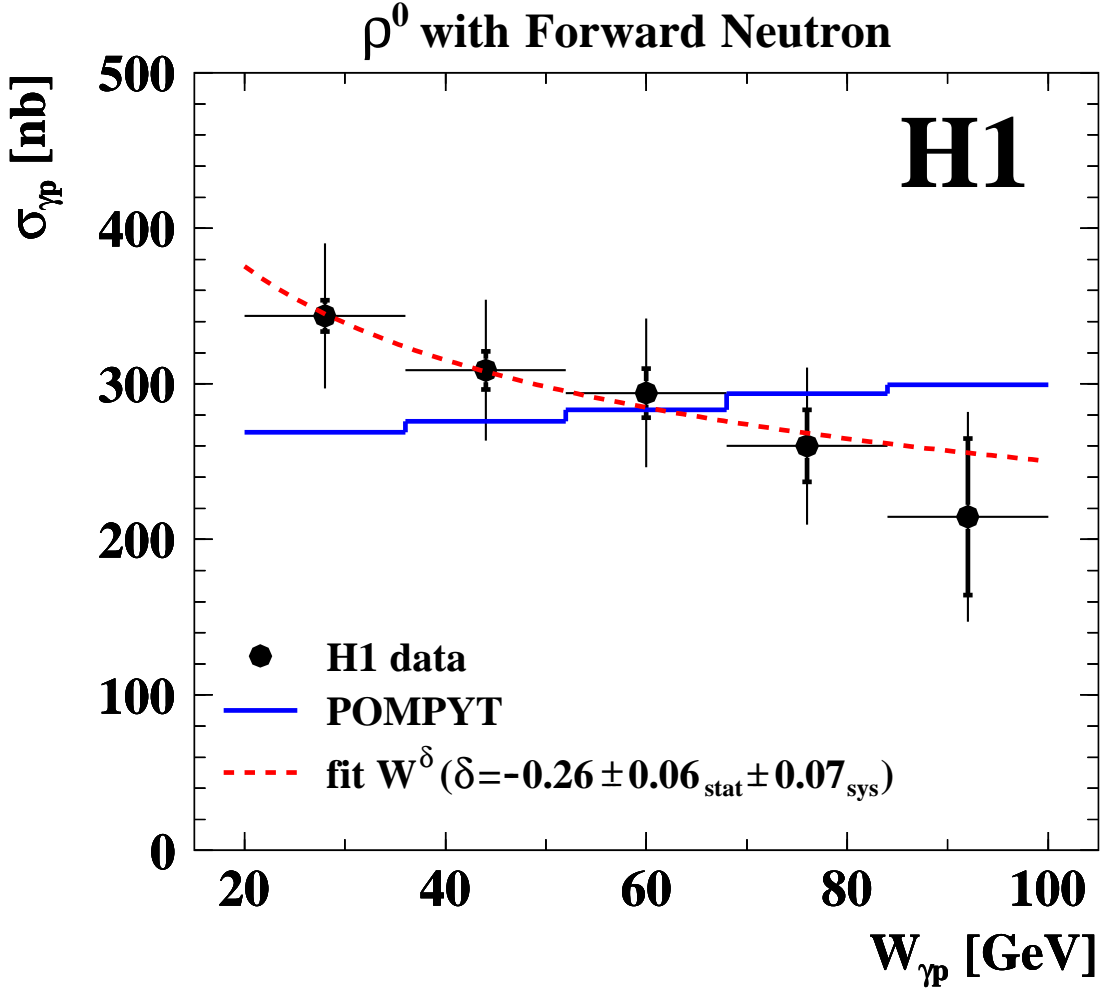}
 \setlength{\unitlength}{1cm}
\caption{Cross section of the reaction $\gprho$ as
         a function of $\Wgp$ compared to the prediction 
         from POMPYT MC program, which is normalised to the data.
         The dashed curve represents the Regge motivated fit
         $\sigma \propto W^{\delta}$
         with $\delta = -0.26 \pm 0.06_{\rm stat}\pm 0.07_{\rm sys}$.
         The data points are shown with statistical (inner error bars) and
         total uncertainties (outer error bars) excluding an overall
         normalisation error of $4.4\%$.}
\label{fig:wgp}
\end{figure}

\begin{figure}[hhh]
\center
 \epsfig{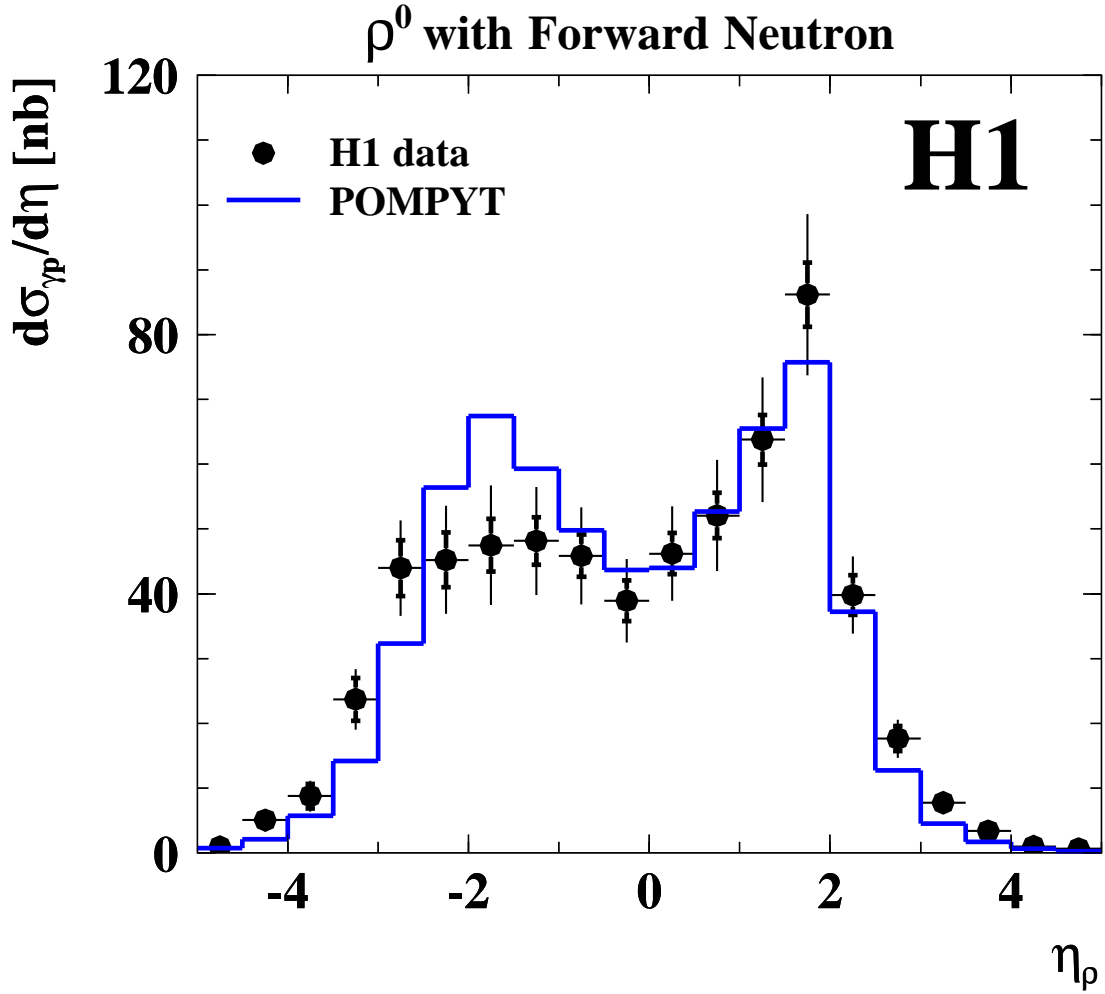}
 \setlength{\unitlength}{1cm}
\caption{The differential cross section d$\sigma_{\gamma p}/{\rm d}\eta$
         as a function of pseudorapidity of the $\rho^0$ meson
         compared to the prediction from the POMPYT MC program,
         which is normalised to the data.
         The inner error bars represent statistical errors and the outer
         error bars are total errors excluding an overall normalisation
         uncertainty of $4.4\%$.}
\label{fig:eta}
\end{figure}

The energy dependence of the reaction $\gprho$ is presented 
in table~\ref{tab:table5} and in figure~\ref{fig:wgp}.
The cross section drops with $\Wgp$ in contrast to
the POMPYT MC expectation, 
where the energy dependence is driven by Pomeron exchange alone. 
A Regge motivated power law fit to the data,
$\sgp(\Wgp) \propto \Wgp^{\delta}$,
yields $\delta = -0.26 \pm 0.06_{stat} \pm 0.07_{sys}$.  
%
The difference in the energy dependence in data and MC is also reflected
in the pseudorapidity distribution of the $\rho^0$ meson,
which is given in table~\ref{tab:table6} and shown in figure~\ref{fig:eta}.

\begin{figure}[hhh]
\center
 \epsfig{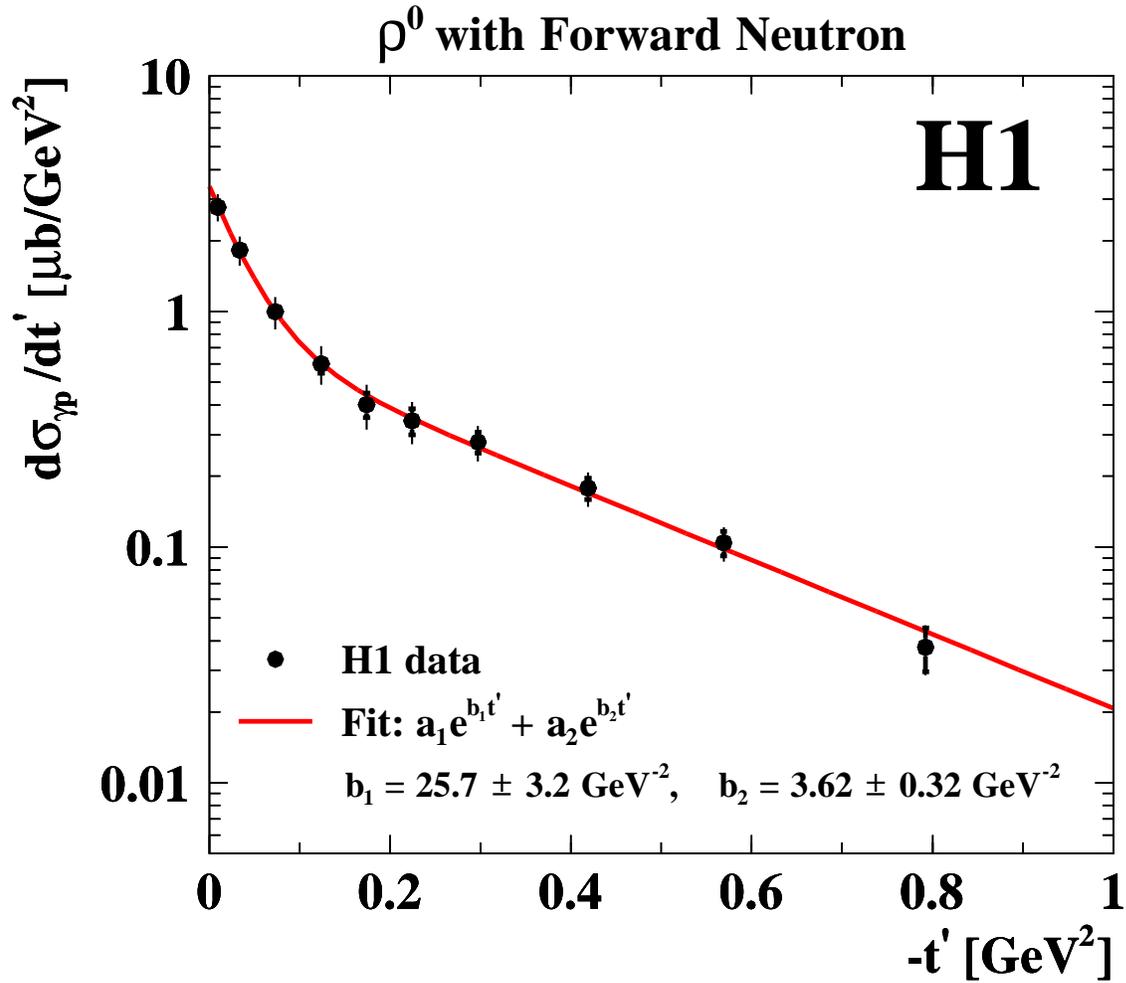}
 \setlength{\unitlength}{1cm}
\caption{Differential cross section d$\sigma_{\gamma p}/{\rm d}t^{\prime}$
         of $\rho^0$ mesons fitted with the sum of two exponential functions.
         The inner error bars represent statistical and uncorrelated
         systematic uncertainties added in quadrature
         and the outer error bars are total uncertainties, 
         excluding an overall normalisation error of $4.4\%$.}
\label{fig:pt2}
\end{figure}

%
Finally, the cross section as a function of the four-momentum 
transfer squared of the $\rho^0$ meson, $t^{\prime}$, 
is given in table~\ref{tab:table7} 
and presented in figure~\ref{fig:pt2}.
It exhibits the very pronounced feature of a strongly changing slope
between the low-$t^{\prime}$ and the high-$t^{\prime}$ regions.
The fit is performed to the sum of two exponential functions:
%
\beq
      \frac{{\rm d}\sgp}{{\rm d}t^{\prime}} = 
            a_1 e^{b_1 t^{\prime}} + a_2 e^{b_2 t^{\prime}}
 \label{eq:fit}
\eeq
%
and yields the following slope parameters: 
%
\beq
     b_1 = (25.72 \pm 3.22_{unc} \pm 0.26_{cor})~{\rm GeV}^{-2}; ~~~
     b_2 = (3.62 \pm 0.30_{unc} \pm 0.10_{cor})~{\rm GeV}^{-2}
 \label{eq:slopes}
\eeq
%
where the first errors include statistical and uncorrelated systematic 
uncertainties and the second errors are due to correlated systematic 
uncertainties.
In a geometric picture, the large value of $b_1$ suggests that 
for a significant part of the data $\rho^0$ mesons are produced
at large impact parameter values of order 
$\langle r^2\rangle = 2b_1\!\cdot\!(\hbar c)^2 \simeq 2 {\rm fm}^2
                      \approx (1.6 R_{\rm p})^2$. 
In other words, photons find pions in a cloud which extends far beyond 
the proton radius.
The small value of $b_2$ corresponds to a target size of $\sim\!0.5$ fm. 
In the DPP interpretation~\cite{Zotov,interf_1,interf_2} 
the observed behaviour is a consequence of the interference
between the amplitudes corresponding to the diagrams {\it a, b} and {\it c}
in figure~\ref{fig:FD},
leading to a slope dependence on the invariant mass 
of the $(n\pi^+)$ system produced at the proton vertex.
Since the forward pion is not detected in this analysis
the $(n\pi^+)$ invariant mass cannot be determined 
with sufficient precision,
which prevents explicit measurement of the $b(m)$ dependence.

\begin{figure}[hhh]
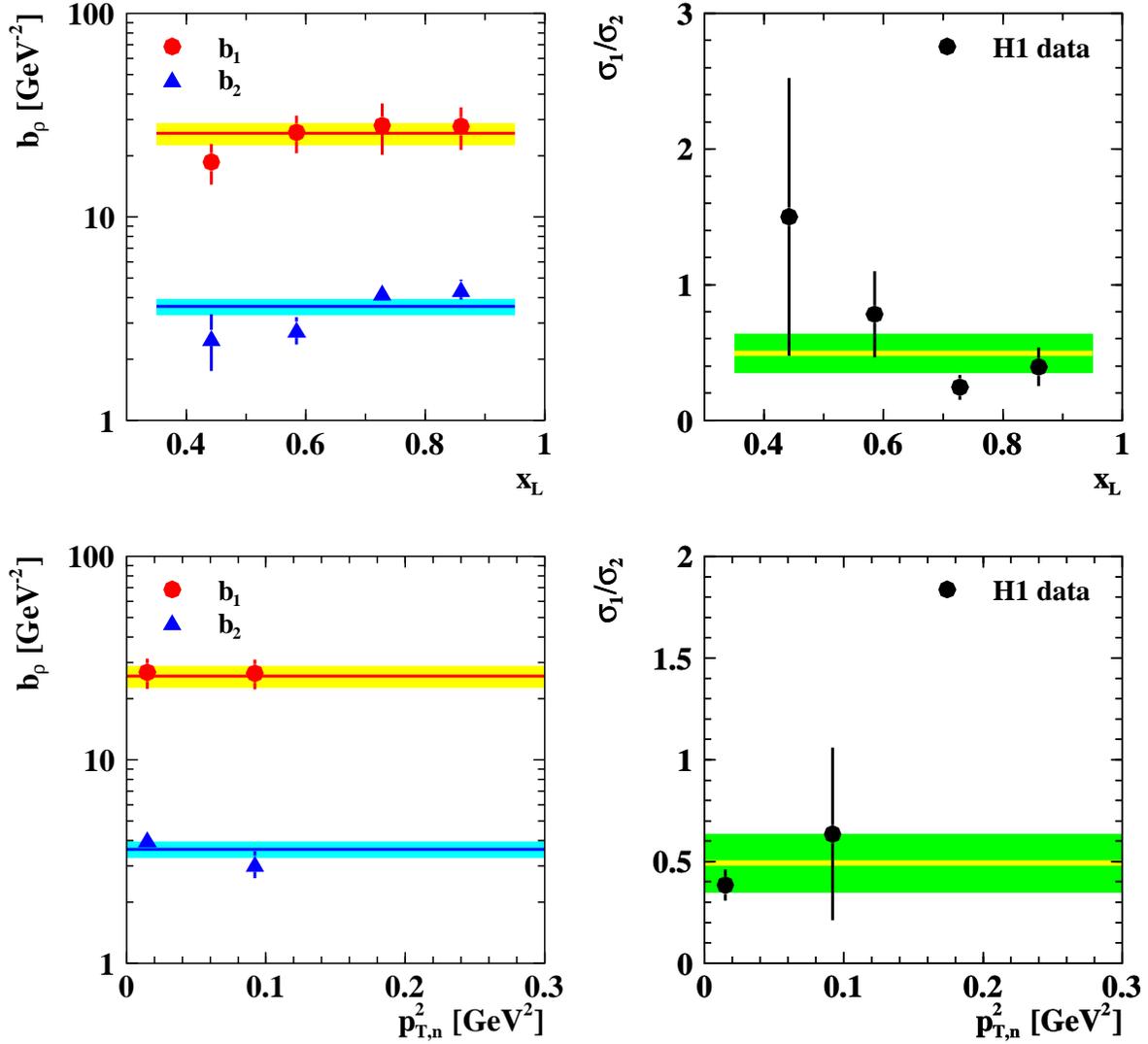

\center
 \epsfig{file=d15-120f12a.eps,width=\textwidth}
 \epsfig{file=d15-120f12b.eps,width=\textwidth}
 \epsfig{file=d15-120f12c.eps,width=\textwidth}
 \setlength{\unitlength}{1cm}
\caption{The two exponential slopes, $b_1$ and $b_2$, obtained by
         fitting the $t^{\prime}$ dependence of the $\rho^0$ mesons (left)
         and the relative contribution of the two exponents
         to the overall cross section of the reaction $\gprho$ (right)
         as a function of $x_L$ (top) and $\ptn^2$ (bottom). 
         The error bars re\-present statistical and systematic uncertainties
         added in quadrature.
         Horizontal lines with error bands show the
         corresponding average values for the full ranges
         of $0.35<x_L<0.95$ and $0<\ptn^2< 0.3~\gevsq$.}
\label{fig:b_rho_xl}
\end{figure}

%
In order to investigate the presence of a possible factorisation 
between the proton and the photon vertices,
the $t^{\prime}$ distribution is studied in bins of $x_L$.
The result of the fit by equation~(\ref{eq:fit}) with $x_L$ dependent 
parameters $a_i(x_L), b_i(x_L)$ is presented in table~\ref{tab:table8}
and in figure~\ref{fig:b_rho_xl} in comparison with the values
given in equation~(\ref{eq:slopes}) for the full $x_L$ range.
Also the evolution with $x_L$ of the ratio of two components, 
$\sigma_1/\sigma_2$,
where $\sigma_i = \frac{a_i}{b_i}(1-e^{-b_i})$, is shown.   
Given the large experimental uncertainties
no strong conclusion about factorisation of the two vertices
can be drawn.

\subsection{$\gpi$ cross section}
\label{sec:res_gpi}

\begin{figure}[ppp]
\center
 \epsfig{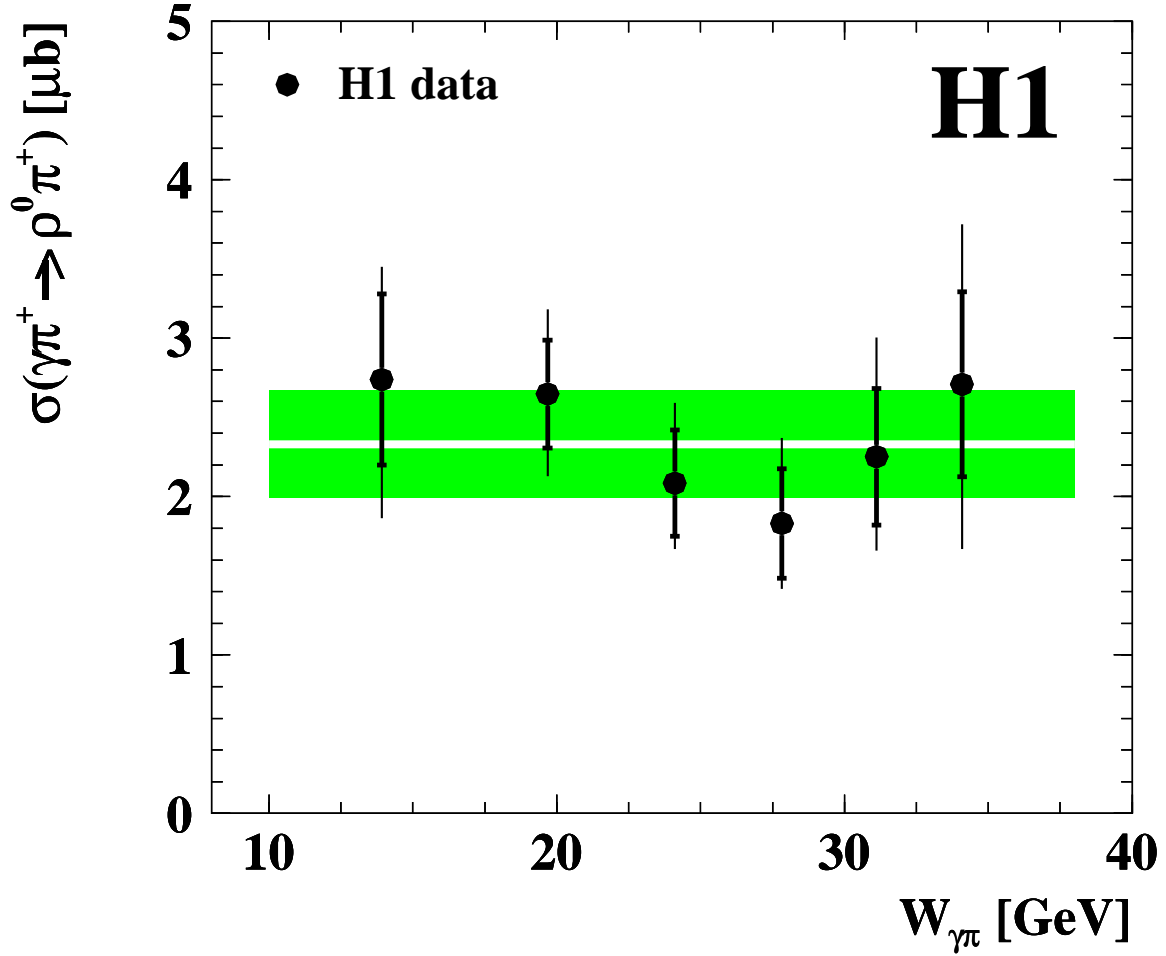}
 \setlength{\unitlength}{1cm}
\caption{Elastic cross section,
         $\sgpi^{\rm el} \equiv \sigma ({\gamma\pi^+} \to \rho^0\pi^+)$,
         extracted in the one-pion-exchange approximation
         as a function of the photon-pion energy, $\Wgpi$.
         The inner error bars represent the total experimental uncertainty
         and the outer error bars are experimental and model uncertainties
         added in quadrature, where the model error is due to
         pion flux uncertainties.
         The dark shaded band represents the average value for the full $\Wgpi$
         range as given in equation~(\ref{eq:sgp3}).}
\label{fig:sigma_gpi}
\end{figure}

The pion flux models compatible with the data in shape of the $x_L$  
distribution are used to extract the photon-pion cross sections 
from $\dsig x_L$ in the OPE approximation. 
The results are presented in table~\ref{tab:table9}
and in figure~\ref{fig:sigma_gpi}.
As a central value the Holtmann flux~\cite{flux_Holt} is used,
and the largest difference to the other three 
predictions~\cite{flux_Bish,flux_KPP,flux_MST}
provides an estimate of the model uncertainty which is $\sim\!19\%$
on average. 
From the total $\gp$ cross section in equation~(\ref{eq:sgp2}) 
and using the pion flux~(\ref{eq:piflux}-\ref{eq:piff}) 
integrated in $x_L$ and $\ptn$, $\Gamma_{\pi} = 0.056$,
the cross section of elastic photoproduction of $\rho^0$ on a pion target
is determined at an average energy $\langle \Wgpi \rangle \simeq 24$ GeV:
%
\beq
 \sigma (\gamma \pi^+\to \rho^0\pi^+) = 
        (2.33\pm0.34 (\rm exp) ^{+0.47}_{-0.40} (\rm model))~\mu\rm b,
 \label{eq:sgp3}
\eeq
%
where the experimental uncertainty includes statistical, systematic and 
normalisation errors added in quadrature, while the model error 
is due to the uncertainty in the pion flux integral 
obtained for the different flux parametrisations
compatible with our data.

Theoretical studies of leading neutron production 
in $ep$ collisions~\cite{flux_Holt,flux_KPP} suggest 
that in addition to the pion exchange process other processes\footnote{For 
      inclusive leading neutron production,
      $\rho, a_2$ trajectories should be considered,
      while for the exclusive reaction~(\ref{eq:reaction}) the diagrams 
      shown in figures~\ref{fig:FD}b,c become important at larger $t$.}  
may contribute at $10-20\%$ level.
To suppress these contributions it is recommended to perform 
cross section measurements in the `OPE safe' phase space region: 
low $\ptn$ and high $x_L$.
In order to investigate a possible influence of non-OPE contributions
the extraction of the photon-pion cross section is repeated 
for two additional regions,
in which the validity of pure OPE  is assumed.
The cross sections for the full FNC acceptance range 
$(\theta_n<0.75~{\rm mrad},~0.35<x_L<0.95)$
and for the OPE2 sub-sample
$(\ptn<200~{\rm MeV},~0.65<x_L<0.95)$ together with the value~(\ref{eq:sgp3})
obtained for the OPE1 sample
are presented in table~\ref{tab:table10}.
The values of $\sigma (\gamma \pi^+\to \rho^0\pi^+)$ 
extracted in these three different phase space regions 
agree well within the experimental errors.
%
%
Thus no evidence for an extra contribution 
beyond the OPE is found in the full FNC acceptance range 
for the exclusive reaction studied here.  
 
Taking a value of $\sigma (\gamma p \to \rho^0 p) = (9.5 \pm 0.5)~\mu{\rm b}$
at the corresponding energy $\langle W \rangle=24$ GeV, 
which is an interpolation between fixed target and HERA measurements 
(see e.g. figure~10 in \cite{ZEUS_rho}), one obtains for the ratio
$r_{\rm el} = \sigma_{\rm el}^{\gamma\pi}/\sigma_{\rm el}^{\gamma p} = 0.25 \pm 0.06$.
A similar ratio, but for the total cross sections at $\langle W \rangle=107$ GeV,
has been estimated by the ZEUS collaboration as
$r_{\rm tot} = \sigma_{\rm tot}^{\gamma\pi}/\sigma_{\rm tot}^{\gamma p} 
= 0.32 \pm 0.03$~\cite{Chekanov:2002}. 
Both ratios are significantly smaller than
their respective expectations, based on simple considerations.
For $r_{\rm tot}$, a value of $2/3$ is predicted by the 
additive quark model~\cite{aqm}, while
      $r_{\rm el} = (\frac{b_{\gp}}{b_{\gpi}}) \cdot 
      (\sigma_{\rm tot}^{\gpi}/\sigma_{\rm tot}^{\gp})^2 = 0.57 \pm 0.03$ 
can be deduced by combining the optical theorem,
the eikonal approach~\cite{eikonal_orig} relating cross sections 
with elastic slope parameters~\cite{Povh87}
and the data on $pp, \pi^+p$~\cite{data_slopes} and $\gp$~\cite{ZEUS_rho} 
elastic scattering.
Such a suppression of the cross section 
is usually attributed to rescattering, 
or absorptive corrections~\cite{abs_corr_0,abs_corr_1,abs_corr_2,abs_corr_3}, 
which are essential for leading neutron production.
For the exclusive reaction $\gprho$ studied here this would imply 
an absorption factor of $K_{abs} = 0.44 \pm 0.11$.
It is interesting to note, that this value is similar 
to the somewhat different, but conceptually related damping factor
in diffractive dijet photoproduction,
the rapidity gap survival probability, $\langle S^2 \rangle \simeq 0.5$, 
which has been determined by the H1 collaboration~\cite{H1_LRG_2j1,H1_LRG_2j2,H1_vfps_2j}.

\section{Summary}

The photoproduction cross section for exclusive $\rho^0$ production associated with
a leading neutron is measured for the first time at HERA.
The integrated $\gamma p$ cross section in the kinematic range
$20<\Wgp<100$ GeV, $0.35<x_L<0.95$ and $\theta_n<0.75$ mrad 
is determined with $2\%$ statistical and $14.6\%$ systematic precision.
The elastic photon-pion cross section, 
$\sigma(\gamma\pi^+ \to \rho^0\pi^+)$, at $\langle \Wgpi \rangle = 24$ GeV
is extracted in the one-pion-exchange approximation.

Single and double differential $\gamma p$ cross sections are measured.
The differential cross section d$\sigma/{\rm d}t^{\prime}$ shows
a behaviour typical for exclusive double peripheral exchange processes.

The differential cross sections for the leading neutron are sensitive to the
pion flux models. While the shape of the $x_L$ distribution is well reproduced 
by most of the pion flux parametrisations, the $x_L$ dependence of the $p_T$ slope
of the leading neutron is not described by any of the existing models.
This may indicate that the proton vertex factorisation hypothesis does not hold
in exclusive photoproduction, e.g. due to large absorptive effects
which are expected to play an essential r\^ole in soft peripheral processes. 
The estimated cross section ratio for the elastic photoproduction 
of $\rho^0$ mesons on the pion and on the proton, 
$r_{\rm el} = \sigma_{\rm el}^{\gamma\pi}/\sigma_{\rm el}^{\gamma p} = 0.25 \pm 0.06$, 
suggests large absorption corrections, of the order of $60\%$, suppressing
the rate of the studied reaction $\gprho$.

\section*{Acknowledgements}

We are grateful to the HERA machine group whose outstanding
efforts have made this experiment possible. 
We thank the engineers and technicians for their work in constructing and
maintaining the H1 detector, our funding agencies for financial support, 
the DESY technical staff for continual assistance
and the DESY directorate for support and for the hospitality 
which they extend to the non DESY members of the collaboration.
We would like to give credit to all partners contributing to the EGI 
computing infrastructure for their support for the H1 Collaboration.




\clearpage



\begin{table}[tb]
\centering
{
\begin{tabular}{|c||c|r|r|r|r||c|r|r|r|r|}
 \hline
                                                         &
 \multicolumn{5}{|c||}{($\ptn < x_L \cdot 0.69$ GeV)}    &  
 \multicolumn{5}{|c|}{($\ptn < 0.2$ GeV)}               \\ 
 \cline{2-11}          
 \raisebox{-.1cm}{\large $x_L$}                          &  
 \raisebox{-.1cm}{${\rm d}\sgp/{\rm d}x_L$}              & 
 \raisebox{-.1cm}{$\delta_{stat}$}                       &
 \raisebox{-.1cm}{$\delta_{sys}^{unc}$}                  &
 \raisebox{-.1cm}{$\delta_{sys}^{cor}$}                  &
 \raisebox{-.1cm}{$\delta_{tot}$}                        &
 \raisebox{-.1cm}{${\rm d}\sgp/{\rm d}x_L$}              & 
 \raisebox{-.1cm}{$\delta_{stat}$}                       &
 \raisebox{-.1cm}{$\delta_{sys}^{unc}$}                  &
 \raisebox{-.1cm}{$\delta_{sys}^{cor}$}                  &
 \raisebox{-.1cm}{$\delta_{tot}$}                       \\ 
                                                         &
 \raisebox{0.02cm}{$[\mu{\rm b}]$}                       &  
 \raisebox{+.02cm}{$[\%]$} & \raisebox{+.02cm}{$[\%]$}   &
 \raisebox{+.02cm}{$[\%]$} & \raisebox{+.02cm}{$[\%]$}   &
 \raisebox{+.02cm}{$[\mu{\rm b}]$}                       &  
 \raisebox{+.02cm}{$[\%]$} & \raisebox{+.02cm}{$[\%]$}   &
 \raisebox{+.02cm}{$[\%]$} & \raisebox{+.02cm}{$[\%]$}  \\
 \hline 
 & & & & & & & & & & \\ [-.40cm]
 $0.35-0.45$ & $0.213$ & $9.8$ & $10.6$ & $15.1$ & $20.9$ & $0.119$ & $11.2$ & $10.3$ & $15.2$ & $21.5$ \\
 $0.45-0.55$ & $0.398$ & $7.0$ & $ 9.8$ & $15.4$ & $19.5$ & $0.164$ & $ 8.6$ & $ 7.5$ & $15.3$ & $19.1$ \\ 
 $0.55-0.65$ & $0.530$ & $5.9$ & $ 7.2$ & $15.7$ & $18.2$ & $0.190$ & $ 7.6$ & $ 7.8$ & $15.4$ & $18.9$ \\
 $0.65-0.75$ & $0.761$ & $4.1$ & $ 6.9$ & $12.8$ & $15.1$ & $0.274$ & $ 5.1$ & $ 9.5$ & $12.0$ & $16.2$ \\  
 $0.75-0.85$ & $0.806$ & $3.6$ & $ 5.0$ & $11.7$ & $13.2$ & $0.354$ & $ 4.1$ & $ 5.8$ & $10.7$ & $12.8$ \\
 $0.85-0.95$ & $0.402$ & $5.4$ & $19.4$ & $12.8$ & $23.9$ & $0.204$ & $ 6.3$ & $15.0$ & $11.2$ & $19.7$ \\
 [+.10cm] \hline
\end{tabular}

}
\caption{
 Differential photoproduction cross sections ${\rm d}\sgp/{\rm d}x_L$ 
 for the exclusive process $\gamma p \to \rho^0 n \pi^+$ 
 in two regions of neutron transverse momentum and $20\!<\!\Wgp\!<\!100$ GeV.
 The statistical, uncorrelated and correlated systematic uncertainties,
 $\delta_{stat}$, $\delta_{sys}^{unc}$ and $\delta_{sys}^{cor}$ respectively,
 are given together with the total uncertainty $\delta_{tot}$, which does not
 include the global normalisation error of $4.4\%$.
%
}
\label{tab:table2}
\end{table}


\begin{table}[tb]
\centering
{
\begin{tabular}{|c|c|c|c||c|r|r|r|r|}
 \hline
 \raisebox{-.40cm}{$x_L$ range}                        & 
 \raisebox{-.40cm}{$\langle x_L \rangle$}              &  
 \raisebox{-.10cm}{$\ptn^2$ range}                     &
 \raisebox{-.15cm}{$\langle \ptn^2 \rangle$}           &  
 \raisebox{-.15cm}{\large $\frac{{\rm d^2}\sigma_{\gamma p}}
                   {{\rm d}x_L{\rm d}\ptn^2}$}         &       
 \raisebox{-.1cm}{$\delta_{stat}$}                     &
 \raisebox{-.1cm}{$\delta_{sys}^{unc}$}                &  
 \raisebox{-.1cm}{$\delta_{sys}^{cor}$}                &
 \raisebox{-.1cm}{$\delta_{tot}$}                     \\ 
 &  & \raisebox{+.10cm}{[GeV$^2$]}                     & 
      \raisebox{+.10cm}{[GeV$^2$]}                     &
      \raisebox{+.10cm}{ $[\mu{\rm b}/\GeV^2]$}        & 
 \raisebox{+.10cm}{$[\%]$} & \raisebox{+.10cm}{$[\%]$} &
 \raisebox{+.10cm}{$[\%]$} & \raisebox{+.10cm}{$[\%]$} \\
 \hline
 & & & & & & & & \\ [-.40cm] 
 $0.35-0.50$ & $0.440$ & $0.00-0.01$ & $0.00499$ & $3.178$ & $13.9$ & $6.3$ & $14.8$ & $21.3$ \\
             &         & $0.01-0.03$ & $0.01998$ & $3.545$ & $12.1$ & $5.4$ & $12.7$ & $18.4$ \\ 
             &         & $0.03-0.06$ & $0.04495$ & $2.974$ & $13.7$ & $6.1$ & $12.7$ & $19.7$ \\ 
 \hline 
 $0.50-0.65$ & $0.581$ & $0.00-0.01$ & $0.00492$ & $5.242$ & $10.5$ & $4.3$ & $14.0$ & $18.0$ \\
             &         & $0.01-0.03$ & $0.01969$ & $4.925$ & $ 8.6$ & $4.1$ & $12.9$ & $16.0$ \\ 
             &         & $0.03-0.06$ & $0.04429$ & $3.344$ & $11.7$ & $4.7$ & $13.9$ & $18.8$ \\ 
             &         & $0.06-0.12$ & $0.08719$ & $2.775$ & $11.2$ & $7.3$ & $13.7$ & $19.1$ \\ 
 \hline 
 $0.65-0.80$ & $0.728$ & $0.00-0.01$ & $0.00489$ & $9.623$ & $ 6.3$ & $4.5$ & $11.4$ & $13.8$ \\
             &         & $0.01-0.03$ & $0.01957$ & $7.229$ & $ 5.5$ & $5.5$ & $12.0$ & $14.3$ \\ 
             &         & $0.03-0.06$ & $0.04403$ & $5.333$ & $ 7.3$ & $5.7$ & $12.2$ & $15.3$ \\ 
             &         & $0.06-0.12$ & $0.08617$ & $2.927$ & $ 8.4$ & $4.8$ & $13.7$ & $16.8$ \\ 
             &         & $0.12-0.20$ & $0.15324$ & $1.494$ & $14.7$ & $6.3$ & $17.9$ & $24.0$ \\ 
 \hline 
 $0.80-0.95$ & $0.863$ & $0.00-0.01$ & $0.00484$ & $7.990$ & $ 7.6$ & $8.5$ & $11.2$ & $16.0$ \\
             &         & $0.01-0.03$ & $0.01935$ & $6.457$ & $ 5.7$ & $7.1$ & $10.9$ & $14.2$ \\ 
             &         & $0.03-0.06$ & $0.04354$ & $3.850$ & $ 7.9$ & $7.4$ & $12.3$ & $16.4$ \\ 
             &         & $0.06-0.12$ & $0.08425$ & $1.580$ & $11.3$ & $7.8$ & $15.7$ & $20.8$ \\ 
             &         & $0.12-0.30$ & $0.16558$ & $0.520$ & $14.1$ & $9.3$ & $18.7$ & $25.2$ \\ 
 [+.10cm] \hline
\end{tabular}

}
\caption{
 Double differential photoproduction cross sections $\d2sxp$
 in the range $20\!<\!\Wgp\!<\!100$ GeV.
 The statistical, uncorrelated and correlated systematic uncertainties,
 $\delta_{stat}$, $\delta_{sys}^{unc}$ and $\delta_{sys}^{cor}$ respectively,
 are given together with the total uncertainty $\delta_{tot}$, which does not
 include the global normalisation error of $4.4\%$.
}
\label{tab:table3}
\end{table}


\begin{table}[htb]
\centering
{
\begin{tabular}{|r|r||c|}
 \hline
  & &  \\ [-0.3cm]
 $x_L$ range  & $\langle x_L \rangle$ & $b_n$ [GeV$^{-2}$] \\
  & &  \\ [-0.3cm]
 \hline
  & &  \\ [-0.4cm]
      $0.35-0.50$ & $0.440$ & $~~2.23 \pm 4.57 \pm 2.10$ \\ 
      $0.50-0.65$ & $0.581$ & $~~8.51 \pm 1.74 \pm 1.10$ \\ 
      $0.65-0.80$ & $0.728$ &  $13.17 \pm 0.90 \pm 0.65$ \\ 
      $0.80-0.95$ & $0.863$ &  $18.21 \pm 0.94 \pm 1.05$ \\ 
\hline
\end{tabular}

}
\caption{
 The effective exponential slope, $b_n$, obtained from the fit
 of double differential photoproduction cross sections $\d2sxp$
 to a single exponential function in bins of $x_L$. 
 The first uncertainty represents the fit error from the statistical 
 and uncorrelated systematic uncertainty and the second one is due to 
 the correlated systematic uncertainty.
}
\label{tab:table4}
\end{table}


\begin{table}[htb]
\centering
{
\begin{tabular}{|c|c|c|}
 \hline
  & &  \\ [-0.3cm]
  $\Wgp$ [GeV] & $\Phi_{\gamma}$ & $\sigma (\gprho)$ [nb]  \\
  & &  \\ [-0.3cm]
 \hline
  & &  \\ [-0.3cm]
      $20-36$  & $0.06306$  &  $343.7 \pm 10.1 \pm 45.4$ \\ 
      $36-52$  & $0.03578$  &  $308.7 \pm 12.3 \pm 43.5$ \\ 
      $52-68$  & $0.02413$  &  $294.2 \pm 15.8 \pm 45.2$ \\ 
      $68-84$  & $0.01769$  &  $260.0 \pm 23.1 \pm 44.9$ \\ 
      $84-100$ & $0.01362$  &  $214.5 \pm 50.2 \pm 45.0$ \\ 
  [+.10cm] 
\hline
\end{tabular}

}
\caption{
 Energy dependence of the exclusive photoproduction of a $\rho^0$ meson 
 associated with a leading neutron, $\gamma p \to \rho^0 n \pi^+$.
 The first uncertainty is statistical and the second is systematic.
 The global normalisation uncertainty of $4.4\%$ is not included.
 $\Phi_{\gamma}$ is the integral of the photon flux~(\ref{eq:gflux})
 in a given $\Wgp$ bin.
}
\label{tab:table5}
\end{table}


\begin{table}[htb]
\centering
{
\begin{tabular}{|c|r||r|r|r|r|}
 \hline
 \raisebox{-.50cm}{\large $\eta_{\rho}$}      &
 \multicolumn{1}{|c||}{\raisebox{-.2cm}{${\rm d}\sgp/{\rm d}\eta$}}  &
 \raisebox{-.2cm}{$\delta_{stat}$}            &
 \raisebox{-.2cm}{$\delta_{sys}^{unc}$}       &
 \raisebox{-.2cm}{$\delta_{sys}^{cor}$}       &
 \raisebox{-.2cm}{$\delta_{tot}$}            \\
                           &
 \multicolumn{1}{|c||}{\raisebox{+.2cm}{[nb]}} &
 \raisebox{+.2cm}{$[\%]$} & 
 \raisebox{+.2cm}{$[\%]$} &
 \raisebox{+.2cm}{$[\%]$} & 
 \raisebox{+.2cm}{$[\%]$} \\
 \hline
 & & & & & \\ [-.40cm]
    $[-5.0;-4.5)$  &  $ 0.9$ & $68. $ & $28. $ & $12. $ & $75.$ \\ 
    $[-4.5;-4.0)$  &  $ 5.1$ & $27. $ & $18. $ & $11. $ & $34.$ \\ 
    $[-4.0;-3.5)$  &  $ 8.8$ & $22. $ & $11. $ & $12. $ & $27.$ \\ 
    $[-3.5;-3.0)$  &  $23.7$ & $14. $ & $ 6.1$ & $12. $ & $20.$ \\ 
    $[-3.0;-2.5)$  &  $44.0$ & $ 9.7$ & $ 4.0$ & $13. $ & $17.$ \\ 
    $[-2.5;-2.0)$  &  $45.2$ & $ 9.3$ & $ 3.1$ & $16. $ & $18.$ \\ 
    $[-2.0;-1.5)$  &  $47.5$ & $ 8.6$ & $ 3.5$ & $17. $ & $19.$ \\ 
    $[-1.5;-1.0)$  &  $48.2$ & $ 7.6$ & $ 2.9$ & $15. $ & $17.$ \\ 
    $[-1.0;-0.5)$  &  $45.9$ & $ 7.1$ & $ 5.9$ & $13. $ & $16.$ \\ 
    $[-0.5;~~~0.0)$ & $38.9$ & $ 8.0$ & $ 3.2$ & $14. $ & $16.$ \\
    $[~~~0.0;+0.5)$ & $46.2$ & $ 6.9$ & $ 5.7$ & $13. $ & $16.$ \\ 
    $[+0.5;+1.0)$  &  $52.1$ & $ 6.7$ & $ 7.1$ & $13. $ & $16.$ \\ 
    $[+1.0;+1.5)$  &  $63.8$ & $ 6.0$ & $ 5.4$ & $13. $ & $15.$ \\ 
    $[+1.5;+2.0)$  &  $86.2$ & $ 5.8$ & $ 4.4$ & $13. $ & $14.$ \\ 
    $[+2.0;+2.5)$  &  $39.8$ & $ 7.7$ & $ 3.1$ & $12. $ & $15.$ \\ 
    $[+2.5;+3.0)$  &  $17.7$ & $11. $ & $ 4.0$ & $12. $ & $17.$ \\ 
    $[+3.0;+3.5)$  &  $ 7.8$ & $17. $ & $ 6.8$ & $12. $ & $22.$ \\ 
    $[+3.5;+4.0)$  &  $ 3.4$ & $26. $ & $11. $ & $12. $ & $30.$ \\ 
    $[+4.0;+4.5)$  &  $ 1.0$ & $55. $ & $21. $ & $11. $ & $60.$ \\ 
    $[+4.5;+5.0)$  &  $ 0.7$ & $64. $ & $33. $ & $11. $ & $73.$ \\ 
  [+.10cm] 
\hline
\end{tabular}

}
\caption{
 Differential photoproduction cross section ${\rm d}\sgp/{\rm d}\eta$
 for the exclusive process $\gamma p \to \rho^0 n \pi^+$ 
 as a function of the $\rho^0$ pseudorapidity in the kinematic range
 $0.35\!<\!x_L\!<\!0.95$, $\theta_n\!<\!0.75$ mrad and $20\!<\!\Wgp\!<\!100$ GeV.
 The statistical, uncorrelated and correlated systematic uncertainties,
 $\delta_{stat}$, $\delta_{sys}^{unc}$ and $\delta_{sys}^{cor}$ respectively,
 are given together with the total uncertainty $\delta_{tot}$, which does not
 include the global normalisation error of $4.4\%$.
%
}
\label{tab:table6}
\end{table}


\begin{table}[htb]
\centering
{
\begin{tabular}{|c|c|c||r|r|r|r|}
 \hline
 \raisebox{-.05cm}{$-t^{\prime}$ range}                & 
 \raisebox{-.05cm}{$\langle -t^{\prime} \rangle$}      &  
 \raisebox{-.05cm}{${\rm d}\sgp/{\rm d}t^{\prime}$}    &       
 \raisebox{-.05cm}{$\delta_{stat}$}                    &
 \raisebox{-.05cm}{$\delta_{sys}^{unc}$}               &  
 \raisebox{-.05cm}{$\delta_{sys}^{cor}$}               &
 \raisebox{-.05cm}{$\delta_{tot}$}                     \\ 
 \raisebox{+.05cm}{[GeV$^2$]}                          & 
 \raisebox{+.05cm}{[GeV$^2$]}                          &
 \raisebox{+.05cm}{$[\mu{\rm b}/\GeV^2]$}              & 
 \raisebox{+.05cm}{$[\%]$} & \raisebox{+.10cm}{$[\%]$} &
 \raisebox{+.05cm}{$[\%]$} & \raisebox{+.10cm}{$[\%]$} \\
 \hline
 & & & & & & \\ [-.40cm] 
     $0.00-0.02$ & $0.0094$ & $2.771$ & $ 4.5$ & $ 2.5$ & $12.1$ & $13.2$ \\
     $0.02-0.05$ & $0.0338$ & $1.821$ & $ 4.9$ & $ 1.7$ & $13.0$ & $14.0$ \\
     $0.05-0.10$ & $0.0727$ & $0.996$ & $ 5.9$ & $ 1.3$ & $14.6$ & $15.8$ \\
     $0.10-0.15$ & $0.1236$ & $0.600$ & $ 8.7$ & $ 1.0$ & $16.3$ & $18.5$ \\
     $0.15-0.20$ & $0.1741$ & $0.402$ & $11.6$ & $ 2.9$ & $17.8$ & $21.4$ \\
     $0.20-0.25$ & $0.2242$ & $0.343$ & $12.0$ & $ 3.7$ & $16.0$ & $20.3$ \\ 
     $0.25-0.35$ & $0.2973$ & $0.279$ & $ 8.6$ & $ 5.1$ & $13.8$ & $17.0$ \\
     $0.35-0.50$ & $0.4189$ & $0.178$ & $ 8.3$ & $ 6.4$ & $12.7$ & $16.4$ \\
     $0.50-0.65$ & $0.5689$ & $0.104$ & $ 9.2$ & $ 7.8$ & $11.6$ & $16.8$ \\
     $0.65-1.00$ & $0.7924$ & $0.037$ & $ 9.4$ & $18.7$ & $11.5$ & $23.9$ \\
 [+.10cm] \hline
\end{tabular}

}
\caption{
 Differential photoproduction cross section ${\rm d}\sgp/{\rm d}t^{\prime}$
 for the exclusive process $\gamma p \to \rho^0 n \pi^+$ 
 as a function of the $\rho^0$ four-momentum transfer squared, $t^{\prime}$,
 in the kinematic range $0.35\!<\!x_L\!<\!0.95$, 
 $\theta_n\!<\!0.75$ mrad and $20\!<\!\Wgp\!<\!100$ GeV.
 The statistical, uncorrelated and correlated systematic uncertainties, 
 $\delta_{stat}$, $\delta_{sys}^{unc}$ and $\delta_{sys}^{cor}$ respectively, 
 are given together with the total uncertainty $\delta_{tot}$, which does not
 include the global normalisation error of $4.4\%$.
}
\label{tab:table7}
\end{table}


\begin{table}[htb]
\centering
{
\begin{tabular}{|c|c||c|c|c|}
 \hline
  & & & & \\ [-0.2cm]
 $x_L$ range  & $\langle x_L \rangle$ & 
 $b_1$ [GeV$^{-2}$] & $b_2$ [GeV$^{-2}$] & $\sigma_1/\sigma_2$\\
  & & & & \\ [-0.3cm]
 \hline
  & & & & \\ [-0.3cm]
      $0.35-0.50$ & $0.440$ &  $18.6 \pm 4.2$ & $2.54 \pm 0.79$ & $1.501 \pm 1.024$ \\ 
      $0.50-0.65$ & $0.581$ &  $26.0 \pm 5.5$ & $2.79 \pm 0.43$ & $0.782 \pm 0.316$ \\ 
      $0.65-0.80$ & $0.728$ &  $28.1 \pm 7.9$ & $4.24 \pm 0.34$ & $0.244 \pm 0.091$ \\ 
      $0.80-0.95$ & $0.863$ &  $27.9 \pm 6.5$ & $4.42 \pm 0.50$ & $0.394 \pm 0.142$ \\ 
[+0.1cm] \hline
  & & & & \\ [-0.3cm]
      $0.35-0.95$ & $0.686$ &  $25.7 \pm 3.2$ & $3.62 \pm 0.32$ & $0.492 \pm 0.143$ \\ 
[+0.1cm] \hline
  & & & & \\ [-0.2cm]
 $\ptn^2$ range [GeV$^2$] & $\langle \ptn^2 \rangle$ [GeV$^2$] & 
 $b_1$ [GeV$^{-2}$] & $b_2$ [GeV$^{-2}$] & $\sigma_1/\sigma_2$\\
  & & & & \\ [-0.3cm]
 \hline
  & & & & \\ [-0.3cm]
      $0.0 -0.04$ & $0.015$  &  $26.8 \pm 4.5$ & $4.07 \pm 0.34$ & $0.384 \pm 0.077$ \\  
      $0.04-0.30$ & $0.092$  &  $26.6 \pm 4.4$ & $3.08 \pm 0.46$ & $0.635 \pm 0.423$ \\ 
[+0.1cm] \hline
\end{tabular}

}
\caption{
 Exponential slopes, $b_1$ and $b_2$, and the ratio $\sigma_1/\sigma_2$,
 obtained from the components of fit~(\ref{eq:fit}) to the
 differential cross section ${\rm d}\sgp/{\rm d}t^{\prime}$ in bins of $x_L$ 
 and in bins of $\ptn^2$.
 The errors represent the statistical and systematic uncertainties
 added in quadrature.
}
\label{tab:table8}
\end{table}


\begin{table}[htb]
\centering
{
\begin{tabular}{|c|c|c|c|}
 \hline
  & & & \\ [-0.3cm]
  $x_L$ range                   &  
  $\Gpi(x_L)$                   &
  $\langle \Wgpi \rangle$ [GeV] &
  $\sigma (\gamma\pi^+ \to \rho^0\pi^+)$ [$\mu$b]  \\
  & & & \\ [-0.3cm]
 \hline
  & & & \\ [-0.3cm]
 $0.35-0.45$ & $0.04407$ & $34.08$ & $2.71 \pm 0.58 ^{+0.82} _{-0.86} $ \\
  & & & \\ [-0.3cm]
 $0.45-0.55$ & $0.07262$ & $31.11$ & $2.25 \pm 0.43 ^{+0.62} _{-0.41} $ \\ 
  & & & \\ [-0.3cm]
 $0.55-0.65$ & $0.10400$ & $27.83$ & $1.83 \pm 0.35 ^{+0.41} _{-0.23} $ \\
  & & & \\ [-0.3cm]
 $0.65-0.75$ & $0.13154$ & $24.10$ & $2.09 \pm 0.34 ^{+0.38} _{-0.25} $ \\  
  & & & \\ [-0.3cm]
 $0.75-0.85$ & $0.13386$ & $19.68$ & $2.65 \pm 0.34 ^{+0.41} _{-0.39} $ \\  
  & & & \\ [-0.3cm]
 $0.85-0.95$ & $0.07431$ & $13.91$ & $2.74 \pm 0.54 ^{+0.46} _{-0.69} $ \\
           [+0.2cm]
\hline
\end{tabular}

}
\caption{
 Energy dependence of elastic $\rho^0$ photoproduction on the pion, 
 $\gamma \pi^+ \to \rho^0 \pi^+$, extracted in the one-pion-exchange 
 approximation using OPE1 sample.
 The first uncertainty represents the full experimental error
 and the second is the model error coming from the pion flux uncertainty 
 (see text).
 $\Gpi(x_L)$ represents the value of the pion flux 
 (\ref{eq:piflux}-\ref{eq:piff}) integrated over the $\ptn<0.2$ GeV range, 
 at a given $x_L$.
}
\label{tab:table9}
\end{table}


\begin{table}[htb]
\centering
{
\begin{tabular}{|c|c|c|c|c|}
 \hline
  & & & & \\ [-0.3cm]
  $x_L$ range                   & 
  $\ptn^{\rm max}$ [GeV]        &  
  $\Gpi     $                   &
  $\langle \Wgpi \rangle$ [GeV] &
  $\sigma (\gamma\pi^+ \to \rho^0\pi^+)$ [$\mu$b]  \\
  & & & & \\ [-0.3cm]
 \hline
  & & & & \\ [-0.3cm]
  $0.35-0.95$ & $x_L\cdot 0.69$ & $0.13815$ & $23.65$ & $2.25 \pm 0.34 ^{+0.54} _{-0.50} $ \\ 
             [+0.2cm]
  $0.35-0.95$ &           $0.2$ & $0.05604$ & $23.65$ & $2.33 \pm 0.34 ^{+0.47} _{-0.40} $ \\ 
             [+0.2cm]
  $0.65-0.95$ &           $0.2$ & $0.03397$ & $19.73$ & $2.45 \pm 0.33 ^{+0.41} _{-0.40} $ \\ 
             [+0.2cm]
\hline
\end{tabular}

}
\caption{
 Cross section of elastic $\rho^0$ photoproduction on the pion, 
 $\gamma \pi^+ \to \rho^0 \pi^+$, extracted in the one-pion-exchange 
 approximation using three different samples: full sample, OPE1 and OPE2. 
 The first uncertainty represents the full experimental error
 and the second is the model error coming from the pion flux uncertainty 
 (see text).
 $\Gpi$ represents the value of the pion flux 
 (\ref{eq:piflux}-\ref{eq:piff}) integrated over the corresponding 
 $(x_L,\ptn)$ range.
}
\label{tab:table10}
\end{table}

\clearpage

\newpage  

\newpage  

\newpage  

\newpage  

\newpage  

\newpage  

\newpage  

\newpage  

\newpage  

\newpage  

\newpage  

\newpage  

\end{document}